\documentclass{jfm}

\usepackage{graphicx}
\usepackage{newtxtext}
\usepackage{newtxmath}
\usepackage{natbib}
\usepackage{hyperref}
\usepackage{comment}
\hypersetup{
    colorlinks = true,
    urlcolor   = blue,
    citecolor  = black,
}

\newcommand{\RomanNumeralCaps}[1]

\title{A few-shot and physically restorable symbolic regression turbulence model based on normalized general effective-viscosity hypothesis}

\author{Ziqi Ji\aff{1} \aff{2}
  ,
  Penghao Duan\aff{3}\corresp{\email{pengduan@hku.hk}}
 \and Gang Du\aff{1}\corresp{\email{dugang@buaa.edu.cn}}}

\affiliation{\aff{1}School of Energy and Power Engineering, Beihang University, Beijing, 100190, China
\aff{2}Department of Mechanical Engineering, City University of Hong Kong, Hong Kong, 999077, China
\aff{3}Department of Mechanical Engineering, University of Hong Kong, Hong Kong 999077, China}

\begin{document}
\maketitle

\begin{abstract}
Turbulence is a complex, irregular flow phenomenon ubiquitous in natural processes and engineering applications. The Reynolds-averaged Navier-Stokes (RANS) method, owing to its low computational cost, has become the primary approach for rapid simulation of engineering turbulence problems. However, the inaccuracy of classical turbulence models constitutes the main drawback of the RANS framework. With the rapid development of data-driven approaches, many data-driven turbulence models have been proposed, yet they still suffer from issues of generalizability and accuracy. In this work, we propose a few-shot, physically restorable, symbolic regression turbulence model based on the normalized general effective-viscosity hypothesis. Few-shot indicates that our model is trained on limited flow configurations spanning only a narrow subset of turbulent flow physics, yet can still outperform the baseline model in substantially different turbulent flows. Physically restorable means our model can nearly revert to the baseline model in regimes satisfying specific physical conditions, using only the symbolic regression training results. The normalized general effective-viscosity hypothesis was proposed in our previous study \citep{ji_enhancing_2026}. Specifically, we first formalize the concept of few-shot data-driven turbulence models. Second, we train our symbolic regression turbulence models using only direct numerical simulation (DNS) data for three-dimensional periodic hill flow slices. Third, we evaluate our models on periodic hill flows, zero pressure gradient flat plate flow, NACA0012 airfoil flows, and NASA Rotor 37 transonic axial compressor flows. One of our symbolic regression turbulence models consistently outperforms the baseline model, and we further demonstrate that this model can nearly revert to baseline behavior in certain flow regimes.
\end{abstract}

\begin{keywords}
Authors should not enter keywords on the manuscript, as these must be chosen by the author during the online submission process and will then be added during the typesetting process (see \href{https://www.cambridge.org/core/journals/journal-of-fluid-mechanics/information/list-of-keywords}{Keyword PDF} for the full list).  Other classifications will be added at the same time.
\end{keywords}

{\bf MSC Codes }  {\it(Optional)} Please enter your MSC Codes here

\section{Introduction}
\label{sec:Introduction}

Turbulence represents an extraordinarily complex and irregular flow phenomenon occurring in liquids, gases, and plasma, rendering it omnipresent in both natural processes and engineering applications \citep{pope2001turbulent}. Nevertheless, conducting expeditious and high-fidelity Computational Fluid Dynamics (CFD) simulations for high Reynolds number flows remains challenging due to the prohibitive computational demands of Large Eddy Simulation (LES) and Direct Numerical Simulation (DNS), coupled with the inherent limitations in accuracy exhibited by conventional turbulence models in Reynolds-averaged Navier-Stokes (RANS) simulations. The progressive advancement of data-driven technologies has facilitated their increasing integration into turbulence modeling frameworks. Consequently, numerous data-driven turbulence models have emerged, demonstrating remarkable efficacy and yielding exceptional results \citep{duraisamy_turbulence_2019, zhang_scientometric_2025}.

For training data-driven turbulence models, numerous recent studies have employed multi-case datasets to enhance model generalizability \citep{shan_modeling_2024,huang_random_2024,sun_development_2024,ho_probabilistic_2024,chen2025feature,cherroud_space-dependent_2025,reissmann_symbolic_2025,zhang_autoturb_2025,wu_development_2024,zhang_three-dimensional_2025,rincon_generalisable_2025}, achieving promising results. However, three principal limitations hinder researchers from leveraging substantially larger datasets for model training. First, computational cost imposes a significant constraint. Larger training datasets typically entail greater computational expense, and incorporating additional flow cases inevitably prolongs the development cycle of data-driven turbulence models, thereby impeding rapid iteration and experimentation. Second, acquiring high-fidelity training data, whether from computational fluid dynamics simulations or experimental measurements, is prohibitively expensive, particularly for industrial-level turbulent flows. This cost barrier prevents researchers from further expanding and refining their training datasets. Third, no well-established scaling law currently exists to quantify how model accuracy improves with increasing training data volume, analogous to those observed in large language models. Furthermore, within the current data-driven turbulence modeling framework, it remains plausible that increasing the quantity of training data yields only marginal improvements in predictive accuracy. Conversely, training data-driven turbulence models on a limited set of flow configurations not only facilitates rapid trial and error but also enables the expeditious generation of individual ``experts'' within the recently prominent ``mixture-of-experts'' framework \citep{cherroud_space-dependent_2025, oulghelou2025machine}. 

In light of these considerations, we propose the concept of few-shot data-driven turbulence models, inspired by the few-shot learning paradigm in machine learning. The formal definition of few-shot data-driven turbulence models is presented in Table \ref{tab: The definition of few-shot data-driven turbulence models.}. The term ``superior performance'' in Table \ref{tab: The definition of few-shot data-driven turbulence models.} indicates that our primary focus is on aggregate model performance rather than requiring the data-driven turbulence model to outperform the baseline model at every spatial location within the flow field. This criterion is motivated by the recognition that different regions of a flow field are inherently interconnected and mutually influential; consequently, it is unrealistic to expect a data-driven turbulence model to surpass the baseline model throughout the entire computational domain. It is important to note that the term ``few-shot'' in the present context differs slightly from its conventional usage in machine learning. In traditional few-shot learning, the emphasis is placed on the limited number of individual data samples, whereas here we emphasize the limited number of flow configurations. This distinction arises because flow fields are theoretically continuous, and the number of training data points for data-driven turbulence models depends primarily on the spatial discretizations of the flow field. Therefore, a more meaningful metric is the number of distinct flow configurations rather than the raw count of data points.

\begin{table}
\begin{tabular}{p{1cm}p{6cm}p{6cm}}
 & 
Definition & 
Example \\ [3pt]
Train & The data-driven turbulence models are trained on a limited set of flow configurations that cover only a narrow subset of the turbulent flow physics. & The model is trained solely on periodic hill flows, which are dominated by geometry-induced separation.\\

Test & The data-driven turbulence models demonstrate overall superior performance compared to baseline turbulence models when applied to flow configurations with substantially different turbulent characteristics beyond the training data. & A data-driven turbulence model trained on low-Reynolds-number, incompressible flows with favorable pressure gradients outperforms the baseline model when applied to the transonic axial compressor NASA Rotor 37, which features high-Reynolds-number, compressible flows, and adverse pressure gradients.
\end{tabular}
\caption{Definition of few-shot data-driven turbulence models.}
\label{tab: The definition of few-shot data-driven turbulence models.}
\end{table}

Many recent studies on data-driven turbulence models have focused on generalizability. One approach to improving such models is to revert to baseline models in regions where classical turbulence models already yield accurate predictions, since the empirical coefficients in these baseline models have been carefully calibrated for simple canonical flows. For example, \citet{wu_development_2024} proposed a conditioned field inversion method to preserve the original calibration of the baseline $k$-$\omega$-SST model within the boundary layer. Although their model demonstrated promising results, the reversion to the baseline behavior was achieved through a shielding function rather than being learned autonomously by the machine learning algorithm. Similarly, \citet{ji_progressive_2026} employed a mixture-of-experts framework to revert to the baseline Spalart–Allmaras (SA) model when simulating wall-bounded turbulence. While this mixture-of-experts approach performed well, it required an additional autoencoder network. From a physical standpoint, the most desirable approach would be for data-driven turbulence models to learn this reversion behavior automatically from the training data, without relying on auxiliary mechanisms or external interventions. However, achieving this level of learning remains a significant challenge for data-driven turbulence modeling.

In this work, we propose a few-shot, physically restorable symbolic regression turbulence model based on the normalized general effective-eddy viscosity hypothesis. Only three-dimensional slices of DNS periodic hill flow data are employed for training. We construct two models incorporating 3 and 5 tensors, respectively; the 3-tensor model (SR 3T) includes only the independent tensors relevant to two-dimensional flows, while the 5-tensor model is designated as SR 5T. In addition to the periodic hill flow, we evaluate the models on a zero pressure gradient flat plate, NACA0012 airfoil flow, and the NASA Rotor 37 transonic axial compressor rotor flow. Overall, the SR 5T model outperforms both the baseline $k$-$\omega$-SST model and the SR 3T model. Furthermore, we observe that most of the additional terms introduced in the SR 5T model physically vanish in the near-wall region for the zero-pressure-gradient flat plate and in the non-wake regime downstream of the NASA Rotor 37 blades, effectively restoring the baseline $k$-$\omega$-SST formulation.

The remainder of this article is organized as follows: $\S$ \ref{sec:Methodology} introduces the methodology employed in this study. $\S$ \ref{sec: Results} presents the results obtained from our analysis. $\S$ \ref{sec:Discussion} discusses the basis tensors utilized in this work. Finally, $\S$ \ref{sec:Conclusion} summarizes the main findings and conclusions.

\section{Methodology}
\label{sec:Methodology}

\subsection{The normalized form of general effective-viscosity hypothesis}
\label{sec:The normalized form of general effective-viscosity hypothesis}

The original formulation of the general effective-viscosity hypothesis, as proposed by \citet{pope_more_1975}, can be expressed as:
\begin{align}
\left\{\begin{array}{l}
\mathsfbi{b}=\sum_{i=1}^w g_i \mathsfbi{T}_i \\
g_i=f\left(I_1 \sim I_a, q_1 \sim q_b\right)
\end{array}\right.,
\label{eq: general effective-viscosity hypothesis}
\end{align}
where $\mathsfbi{b}$ denotes the Reynolds stress anisotropy tensor, defined as $\mathsfbi{b}=\mathsfbi{\tau} / 2k-\mathsfbi{\mathsfbi{I}} / 3$, $\mathsfbi{\tau}$ is Reynolds stress. The terms $\left(I_1, \ldots, I_a\right)$ represent the tensor invariants, while $\left(q_1, \ldots, q_b\right)$ signify additional mean flow characteristics. $\{\mathsfbi{T}_i , i=1, \ldots, w\}$ constitute the tensor basis, which can be derived from the mean flow field results. The coefficients $g_i$ are functions of the tensor invariants $\left(I_1, \ldots, I_a\right)$ and the aforementioned mean flow characteristics $\left(q_1, \ldots, q_b\right)$.

Based on the original version of the general effective-viscosity hypothesis, we have proposed a normalized formulation that is more suitable for data-driven methods and improves generalizability \citep{ji_enhancing_2026}, which can be written as:
\begin{align}
\left\{\begin{array}{l}
\mathsfbi{b}=\sum_{i=1}^w \hat{g}_i \hat{\mathsfbi{T}}_i \\
\hat{g}_i=\hat{f}\left(I_1 \sim I_a, q_1 \sim q_b\right)
\end{array}\right.
\label{eq: gihat},
\end{align}
where $\hat{\mathsfbi{T}}_i=\frac{\mathsfbi{T}_i}{\left||\mathsfbi{T}_i\right||_F}$ denotes the normalized basis tensor, and $\hat{g}_i$ represents the corresponding normalized tensor basis coefficient. 

For the normalized general effective-viscosity hypothesis, all tensors employed satisfy the condition that their Frobenius norm equals unity. Consequently, the relative influence of each term $\hat{g}_i \hat{\mathsfbi{T}}_i$ can be assessed directly by comparing the absolute values of the coefficients $\hat{g}_i$, where a larger absolute value indicates a more significant contribution.

\subsection{Input features and tensor basis}
\label{sec:Input features and tensor basis}

The input features used in this study are divided into two parts: tensor invariants and extra features.

For tensor invariants, we adopt the second-order features \citep{ji_tensor_2024} of the tensor invariants proposed by \citet{wu_physics-informed_2018}, which preserve the original indexing of the tensor invariants and are shown below:
\begin{equation}
\begin{aligned}
\left\{
\begin{aligned}
&I_1 = \operatorname{Tr}(\hat{\mathsfbi{S}}^2) \\
&I_3 = \operatorname{Tr}(\hat{\mathsfbi{R}}^2) \\
&I_4 = \operatorname{Tr}(\hat{\mathsfbi{A}}_p^2) \\
&I_5 = \operatorname{Tr}(\hat{\mathsfbi{A}}_k^2) \\
&I_{15} = \operatorname{Tr}(\hat{\mathsfbi{R}} \hat{\mathsfbi{A}}_p) \\
&I_{16} = \operatorname{Tr}(\hat{\mathsfbi{A}}_p \hat{\mathsfbi{A}}_k) \\
&I_{17} = \operatorname{Tr}(\hat{\mathsfbi{R}} \hat{\mathsfbi{A}}_k) \\
\end{aligned},
\right.
\end{aligned}
\label{eq: I}
\end{equation}
where ``$\operatorname{Tr}$'' denotes the trace of a tensor, and the physical meanings of the tensors are summarized in Table~\ref{Table:input_normalization}. The terms $\hat{\mathsfbi{A}}_p$ and $\hat{\mathsfbi{A}}_k$ are defined as $\hat{\mathsfbi{A}}_p = -\mathsfbi{I} \times \hat{\nabla p}$ and $\hat{\mathsfbi{A}}_k = -\mathsfbi{I} \times \hat{\nabla k}$, respectively. It is worth noting that the normalization factor $\omega \sqrt{k}$ used in Table~\ref{Table:input_normalization} has the dimension of acceleration. This quantity can be interpreted as a characteristic acceleration of turbulent fluctuations, defined as the turbulent velocity scale $\sqrt{k}$ divided by the turbulent time scale $1/\omega$. The normalization method used in Table~\ref{Table:input_normalization} is shown below:
\begin{equation}
\hat{\mathsfbi{\alpha}}=\frac{\mathsfbi{\alpha}}{\|\mathsfbi{\alpha}\|+|\beta|}.
\label{eq: local normalization scheme}
\end{equation}

\begin{table}
  \begin{center}
  \def~{\hphantom{0}}
  \begin{tabular}{p{2cm}p{4.5cm}p{3cm}p{2cm}}
    Normalized raw input $\hat{\mathsfbi{\alpha}}$ &
    Description &
    Raw input $\mathsfbi{\alpha}$ &
    Normalization factor $\beta$ \\[3pt]
    $\hat{\mathsfbi{S}}$ &
    Strain-rate tensor &
    $\mathsfbi{S}$ &
    $\omega$ \\
    $\hat{\mathsfbi{R}}$ &
    Rotation-rate tensor &
    $\mathsfbi{R}$ &
    $\omega$ \\
    $\hat{\nabla p}$ &
    Pressure gradient &
    $\nabla p$ &
    $\omega \sqrt{k}$ \\
    $\hat{\nabla k}$ &
    Turbulence kinetic energy gradient &
    $\nabla k$ &
    $\omega \sqrt{k}$ \\
  \end{tabular}
  \caption{Normalization of strain-rate tensor, rotation-rate tensor, pressure gradient, and turbulence kinetic energy gradient. $\omega$ represents the specific dissipation rate, while $k$ represents the turbulent kinetic energy. The $\mathsfbi{S}=\frac{1}{2}\left(\nabla \mathsfbi{U}+(\nabla \mathsfbi{U})^{\mathrm{T}}\right)$ and $\mathsfbi{R}=\frac{1}{2}\left(\nabla \mathsfbi{U}-(\nabla \mathsfbi{U})^{\mathrm{T}}\right)$ are the strain-rate tensor and rotation-rate tensor, respectively, while $\nabla p, \nabla k$ represent the pressure gradient and turbulence kinetic energy gradient, respectively.}
  \label{Table:input_normalization}
  \end{center}
\end{table}

The additional features are listed in Table \ref{tab:extra_features}. In this study, the dissipation rate $\varepsilon$ is computed as $\epsilon = 0.09 k \omega$. It is worth noting that the molecular viscous stress can be expressed as $\mathsfbi{\tau}^{(\nu)} = -2 \nu \mathsfbi{S}$. Consequently, $q_4$ can be interpreted as the ratio of the Reynolds stress intensity to the molecular viscous stress intensity. The normalization method adopted for the features in Table \ref{tab:extra_features} is described below.
\begin{align}
q_\beta=\frac{\hat{q}_\beta}{\left|\hat{q}_\beta\right|+\left|q_\beta^*\right|}
\label{eq: extra features non-dimensionalization}
\end{align}

\begin{table}
  \begin{center}
\def~{\hphantom{0}}
  \begin{tabular}{p{2cm}p{4cm}p{3cm}p{3cm}}
      Normalized extra features $q_\beta$ & Description & Origin extra features $\hat{q}_\beta$ & Normalization factor $q_\beta^*$ \\[3pt]
      $q_1$ & Ratio of excess rotation rate to strain rate (Q criterion) & $\frac{1}{2}\left(\|\mathsfbi{R}\|^2-\|\mathsfbi{S}\|^2\right)$ & $\|\mathsfbi{S}\|^2$ \\
      $q_2$ & Wall-distance based Reynolds number & $\min \left(\frac{\sqrt{k} d}{50 v}, 2\right)$ & Not applicable \\
      $q_3$ & Ratio of turbulent time scale to mean strain time scale & $\frac{k}{\varepsilon}$ & $\frac{1}{\|\mathsfbi{S}\|}$ \\
      $q_4$ & Ratio of the Reynolds stress intensity to the molecular viscous stress intensity & $k$ & $\nu \|\mathsfbi{S}\|$ \\
      $q_5$ & Ratio of turbulent time scale to mean strain time scale & $\|\mathsfbi{S}\|$ & $\omega$ \\
  \end{tabular}
  \caption{List of extra features.}
  \label{tab:extra_features}
  \end{center}
\end{table}

The tensor basis employed in this study consists of five basis tensors, which are a subset of the tensors proposed by \citet{pope_more_1975}. For consistency with previous studies, we retain the original indexing of the basis tensors. The unnormalized basis tensors are given below:
\begin{equation}
\left\{
\begin{array}{l}
\mathsfbi{T}_1=\hat{\mathsfbi{S}}-\dfrac{1}{3}\operatorname{Tr}\!\left(\hat{\mathsfbi{S}}\right)\mathsfbi{I} \\
\mathsfbi{T}_2=\hat{\mathsfbi{S}}\hat{\mathsfbi{R}}-\hat{\mathsfbi{R}}\hat{\mathsfbi{S}} \\
\mathsfbi{T}_3=\hat{\mathsfbi{S}}^2-\dfrac{1}{3}\operatorname{Tr}\!\left(\hat{\mathsfbi{S}}^2\right)\mathsfbi{I} \\
\mathsfbi{T}_4=\hat{\mathsfbi{R}}^2-\dfrac{1}{3}\operatorname{Tr}\!\left(\hat{\mathsfbi{R}}^2\right)\mathsfbi{I} \\
\mathsfbi{T}_6=\hat{\mathsfbi{R}}^2\hat{\mathsfbi{S}}+\hat{\mathsfbi{S}}\hat{\mathsfbi{R}}^2
-\dfrac{2}{3}\operatorname{Tr}\!\left(\hat{\mathsfbi{S}}\hat{\mathsfbi{R}}^2\right)\mathsfbi{I}
\end{array}
\right..
\label{eq:tensor_basis}
\end{equation}
Based on Eq. (\ref{eq: gihat}), these basis tensors need to be normalized as
$\hat{\mathsfbi{T}}_i=\frac{\mathsfbi{T}_i}{\left||\mathsfbi{T}_i\right||_F}$.
The reason we do not employ $\mathsfbi{T}_5$ in this study is that, according to the periodic hill flow DNS dataset, $\mathsfbi{T}_5$ exhibits pronounced discontinuities and lacks smoothness \citep{ji_enhancing_2026}.

\subsection{Baseline model}
\label{sec: Baseline model}

The governing equations for the origin $k$-$\omega$-SST model are:
\begin{equation}
\left\{
\begin{aligned}
\frac{\partial (\rho k)}{\partial t} + \nabla \cdot (\rho \mathsfbi{U} k)
&= P_k - \beta^* \rho \omega k
+ \nabla \cdot \left[ (\mu + \sigma_k \mu_t) \nabla k \right]
+ S_k, \\[2mm]
\frac{\partial (\rho \omega)}{\partial t} + \nabla \cdot (\rho \mathsfbi{U} \omega)
&= \gamma \frac{P_k}{\nu_t} - \beta \rho \omega^2
+ \nabla \cdot \left[ (\mu + \sigma_{\omega} \mu_t) \nabla \omega \right] \\
&\quad + 2(1 - F_1)\frac{\rho \sigma_{\omega 2}}{\omega}
\nabla k \cdot \nabla \omega
+ S_{\omega}.
\label{eq: kOmegaSST}
\end{aligned}
\right.
\end{equation}
where $P_k$ represents the production of turbulence kinetic energy, which is modeled as:
\begin{align}
P_k = \min\left(\mu_t \nabla \mathsfbi{U} : \nabla \mathsfbi{U}, 10 \beta^* \rho k \omega\right),
\label{eq: Pk}
\end{align}
and $S_k$, $S_{\omega}$ are user-defined source terms.
The turbulence viscosity $\nu_t$ is computed as:
\begin{align}
\nu_t = \frac{a_1 k}{\max(a_1 \omega, S F_2)},
\label{eq: nut}
\end{align}
where $S$ is the magnitude of the mean strain rate and the blending functions $F_1$ and $F_2$ are defined as:
\begin{align}
F_1 &= \tanh\left\{\min\left[\max\left(\frac{\sqrt{k}}{\beta^* \omega y}, \frac{500 \nu}{y^2 \omega}\right), \frac{4 \rho \sigma_{\omega 2} k}{CD_{k\omega} y^2}\right]\right\}^4 \\[1mm]
F_2 &= \tanh\left[\max\left(\frac{2 \sqrt{k}}{\beta^* \omega y}, \frac{500 \nu}{y^2 \omega}\right)\right]^2
\label{eq: F1&F2}
\end{align}
$y$ is the distance to the nearest wall, and $CD_{k\omega}$ is the cross-diffusion term:
\begin{align}
CD_{k w}=\max \left(2 \rho \frac{1}{\sigma_{\omega 2} \omega} \frac{\partial k}{\partial x_i} \frac{\partial \omega}{\partial x_i}, 10^{-10}\right)
\label{eq: CD_{k w}}
\end{align}

The remaining parameters are empirical coefficients defined by \citet{menter_ten_2003}. Finally, according to the Boussinesq linear eddy viscosity hypothesis:
\begin{align}
\mathsfbi{\tau}=\frac{2}{3} k \mathsfbi{I}-2 v_{\mathrm{t}} \mathsfbi{S}
\label{eq: tau_Bous}
\end{align}

\subsection{\label{sec: Symbolic regression}Symbolic regression}

\begin{figure}
	\centering
	\includegraphics[width=1.0\textwidth]{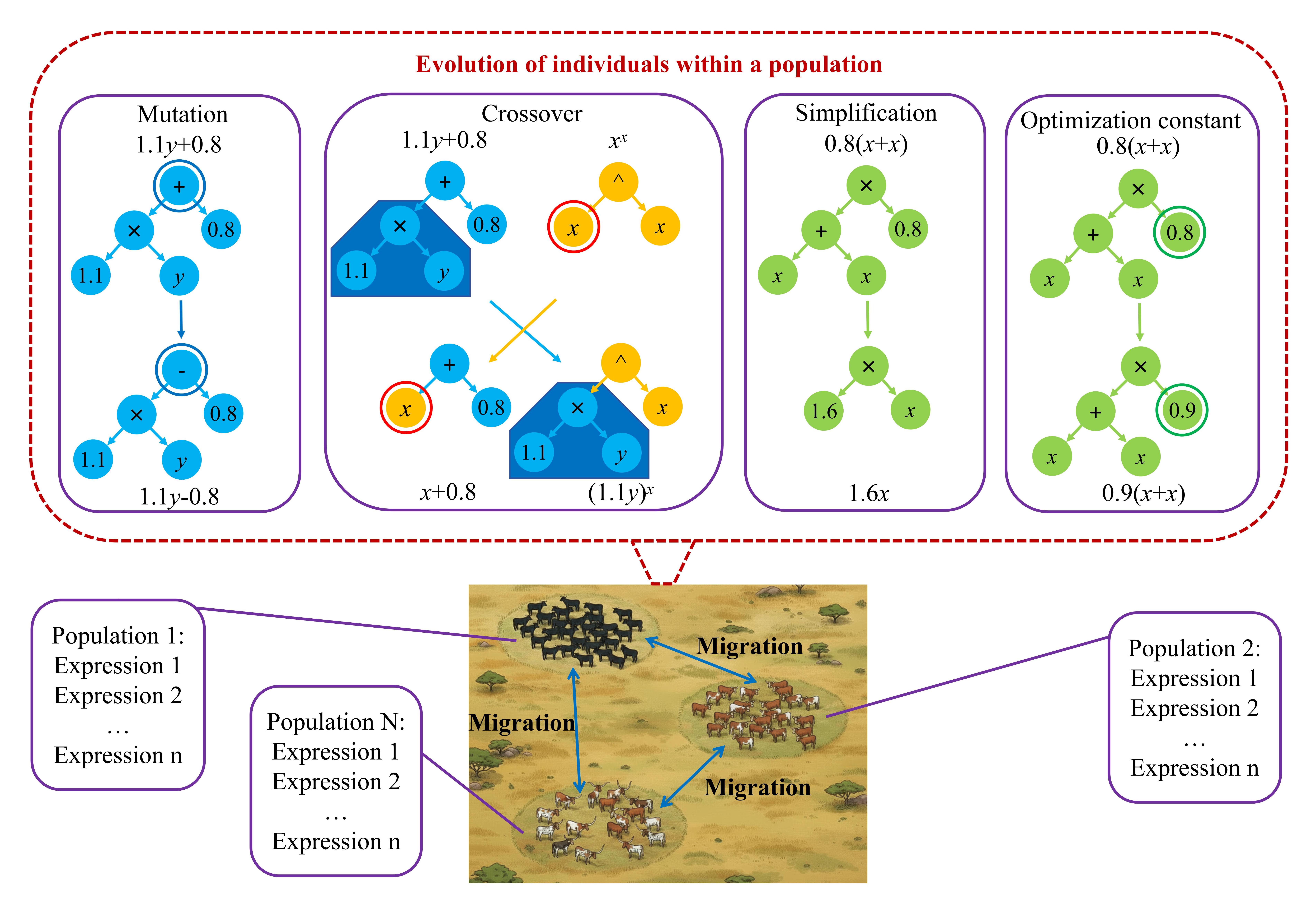}
	\caption{The core algorithmic workflow of PySR. Within each population, expressions evolve continuously through mutation, crossover, simplification, and constant optimization. Concurrently, individuals migrate across different populations.}
	\label{fig:PySR}
\end{figure}

This work uses open-source multi-population evolutionary symbolic regression library PySR \citep{cranmer2023interpretablemachinelearningscience}. As illustrated in Figure~\ref{fig:PySR}, the core idea of the PySR algorithm originates from simulating biological population evolution and leverages binary tree data structures to learn explicit equations. Within each population, expressions undergo continuous evolution through operations such as mutation, crossover, simplification, and constant optimization. Concurrently, individuals are migrating across different populations. The details of the symbolic regression algorithm employed in this work are thoroughly described in the PySR paper~\citep{cranmer2023interpretablemachinelearningscience}.

Within our framework, the symbolic regression is trained to predict normalized tensor basis coefficients $\hat{g_i}$. The regression input comprises tensor invariants, extra features, as elaborated in $\S$ \ref{sec:Input features and tensor basis}. The normalized tensor basis coefficients $\hat{g_i}$ used as training label are gotten by the DNS dataset and following expression \citep{ji_enhancing_2026}:
\begin{align}
\hat{g}_i=\mathsfbi{b}:\mathsfbi{\hat{T}}_i,
\label{eq: gi_normalized}
\end{align}
where ":" denotes the double dot product (tensor contraction).

The key hyperparameters of our symbolic regression algorithm are enumerated in Table \ref{Table: Main hyperparameters of the algorithm}. These hyperparameters are selected primarily based on empirical considerations due to the extensive parameter space of the PySR framework and the prohibitive computational costs associated with training. Consequently, systematic hyperparameter optimization for the symbolic regression implementation employed in this study remains computationally intractable. Our implementation incorporates 8 distinct operators as detailed in Table \ref{Table: Main hyperparameters of the algorithm}. The "neg" operator performs negation, "inv" computes the inverse function, while "square" and "cubic" execute second and third power operations, respectively. Each constant within the formulation is assigned a complexity value of 2. The algorithmic framework employs 16 populations, each containing 100 individuals. During each iteration, we perform 500 mutations for every ten samples extracted from each population, continuing this process for 100 iterations. The number of iterations is 2000. We impose a maximum complexity constraint of 20 for any generated equation, with a maximum depth threshold of 10. Additional hyperparameters conform to the default configurations specified in the PySR library.

\begin{table}
\begin{tabular}{p{8cm} p{5cm}}
Principal parameters of the algorithm & Parameters set-up \\ [3pt]
Operators & +, -, $\times$, $\div$, neg, inv, square, cubic \\ 
Complexity of constants & 2 \\ 
Number of selected features & 10 \\ 
Number of populations & 8 \\ 
Number of individuals in each population & 100 \\ 
Number of total mutations to run, per 10 samples of the population, per iteration & 500 \\ 
Number of iterations & 2000 \\ 
Max complexity of an equation & 20 \\ 
Max depth of an equation & 10 \\ 
\end{tabular}
\caption{\label{Table: Main hyperparameters of the algorithm}Main hyperparameters of the symbolic regression algorithm.}
\end{table}

Furthermore, we implement a symbolic regression strategy to ensure the convergence of results. Specifically, given a set of hyperparameters that we consider sufficient for the symbolic regression algorithm to discover the desired explicit equation (as shown in Table \ref{Table: Main hyperparameters of the algorithm}), we perform $n$ independent runs of the symbolic regression training ($n=5$ in this study). For equations of identical complexity, if the equation with the minimum mean square error (MSE) remains consistent across all $n$ independent runs, we conclude that the result has converged and is independent of the initial conditions of the symbolic regression algorithm.

After obtaining converged results with varying levels of complexity, we select the final equations for posterior computation by balancing accuracy and interpretability. The guiding principle is to ensure that the MSE remains sufficiently low while avoiding excessive complexity that would hinder physical interpretation. We acknowledge that this selection process currently relies on expert judgment and empirical experience; establishing more rigorous, systematic selection criteria remains an important direction for future work.

To ensure that the symbolic regression turbulence models can more readily revert to the baseline model, we employ only the normalized tensor coefficients, except for $\hat{g}_1$, in the symbolic regression training. The term $\hat{g}_1\mathsfbi{\hat{T}}_1$ is fixed as $g_1\mathsfbi{T}_1 = -\frac{v_t}{k}\mathsfbi{T}_1$.

\subsection{Framework}
\label{sec:Framework}

In this study, we employ TCAE as the CFD software for conducting numerical simulations. TCAE, developed by CFD SUPPORT as an extension of the open-source CFD platform OpenFOAM, incorporates a proprietary solver, ``redSolver'', that exhibits exceptional robustness in modeling compressible and transonic flow regimes. In the present investigation, this solver is utilized to simulate the transonic axial compressor rotor, NASA Rotor 37. For incompressible flow cases, we use the ``simpleFoam'' solver, which is part of the standard OpenFOAM distribution. The accuracy of the TCAE code is documented on the TCAE's website.

We employ certain turbulent quantities within the proposed framework, specifically the turbulence kinetic energy $k$ and the specific dissipation rate $\omega$, for normalization purposes, as delineated in Table \ref{Table:input_normalization} and Table \ref{tab:extra_features}. Due to discrepancies between the statistical turbulent quantities computed by DNS and the modeled turbulent quantities derived from RANS, it becomes important to utilize turbulent quantities consistent with the high-fidelity mean flow fields as a normalization factor. We commence by interpolating high-fidelity mean flow data onto the RANS computational grid to acquire compatible turbulent quantities. Subsequently, the interpolated high-fidelity mean flow data are introduced into the $k$ equation and $\omega$ equation. The mean flow is held constant throughout this process, yielding a compatible turbulent flow field. This approach was initially proposed by \citet{weatheritt_novel_2016} and subsequently refined by \citet{schmelzer_discovery_2020} through the introduction of a corrective term in the $k$ equation.

\begin{figure*}
\centering \includegraphics[width=0.8\textwidth]{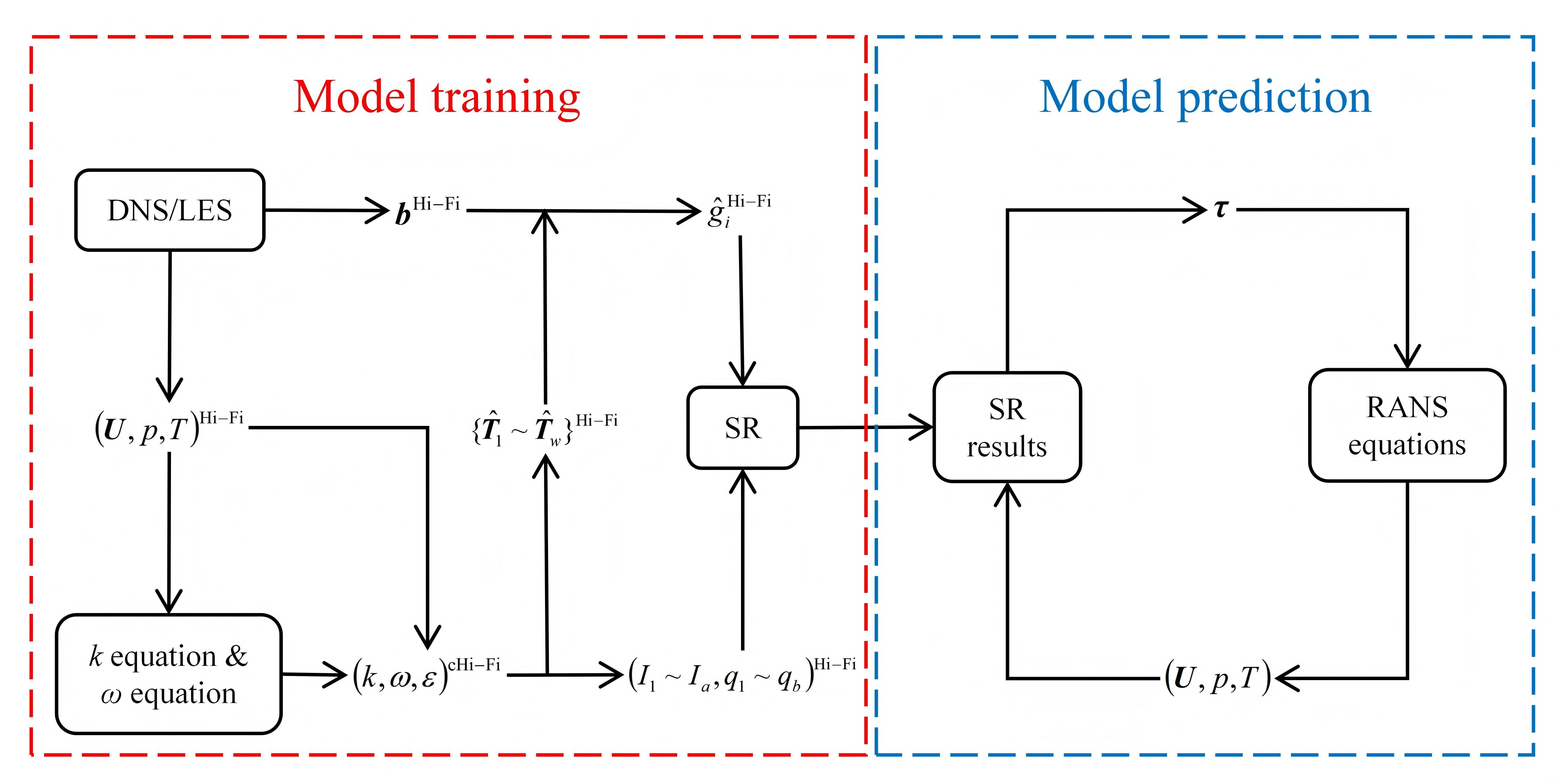} \caption{The framework of our symbolic regression-based turbulence model.}\label{Framework}
\end{figure*}

The methodological framework employed in this investigation is elucidated in detail below. Figure \ref{Framework} provides a comprehensive schematic representation of our proposed approach.

1. Interpolate high-fidelity DNS or LES data onto the RANS mesh to obtain high-fidelity mean flow quantities $(\mathsfbi{U}, p, T)^{\mathrm{Hi-Fi}}$ and Reynolds stress field $\mathsfbi{\tau}^{\mathrm{Hi-Fi}}$.
    
2. Under frozen mean flow conditions, substitute high-fidelity mean flow data $(\mathsfbi{U}, p, T)^{\mathrm{Hi-Fi}}$ into turbulence model scale transport equations ($k$-equation and $\omega$-equation) to generate compatible turbulence fields $(k, \omega,\varepsilon)^{\mathrm{cHi-Fi}}$.
    
3. Construct input features $\left(I_1 \sim I_{\mathrm{a}}, q_1 \sim q_{\mathrm{b}}\right)^{\mathrm{Hi-Fi}}$ and normalized tensor basis $\left\{\hat{\mathsfbi{T}}_1 \sim \hat{\mathsfbi{T}}_w\right\}^{\mathrm{Hi-Fi}}$ based on $(\mathsfbi{U}, p, T)^{\mathrm{Hi-Fi}}$ and $(k, \omega,\varepsilon)^{\mathrm{cHi-Fi}}$. Simultaneously, calculate the normalized tensor basis coefficients $\hat{g}^{\mathrm{Hi-Fi}}$ according to Eq. (\ref{eq: gi_normalized}), using $\mathsfbi{\tau}^{\mathrm{Hi-Fi}}$ and the normalized tensor bases $\left\{\hat{\mathsfbi{T}}_1 \sim \hat{\mathsfbi{T}}_w\right\}^{\mathrm{Hi-Fi}}$.
    
4. Train symbolic regression algorithms using $\left(I_1 \sim I_{\mathrm{a}}, q_1 \sim q_{\mathrm{b}}\right)^{\mathrm{Hi-Fi}}$ as input and $\hat{g}^{\mathrm{Hi-Fi}}_i$ as output.
    
5. Embed the symbolic regression results into the TCAE code. During the numerical solution iteration process, calculate the Reynolds stress tensor $\mathsfbi{\tau}$ using mean flow variables $(\mathsfbi{U}, p, T)$, and incorporate it into the RANS equation system for a coupled solution until the flow field satisfies convergence criteria.

\subsection{Dataset}
\label{sec:Dataset}

\begin{figure*}
\centering \includegraphics[width=0.8\textwidth]{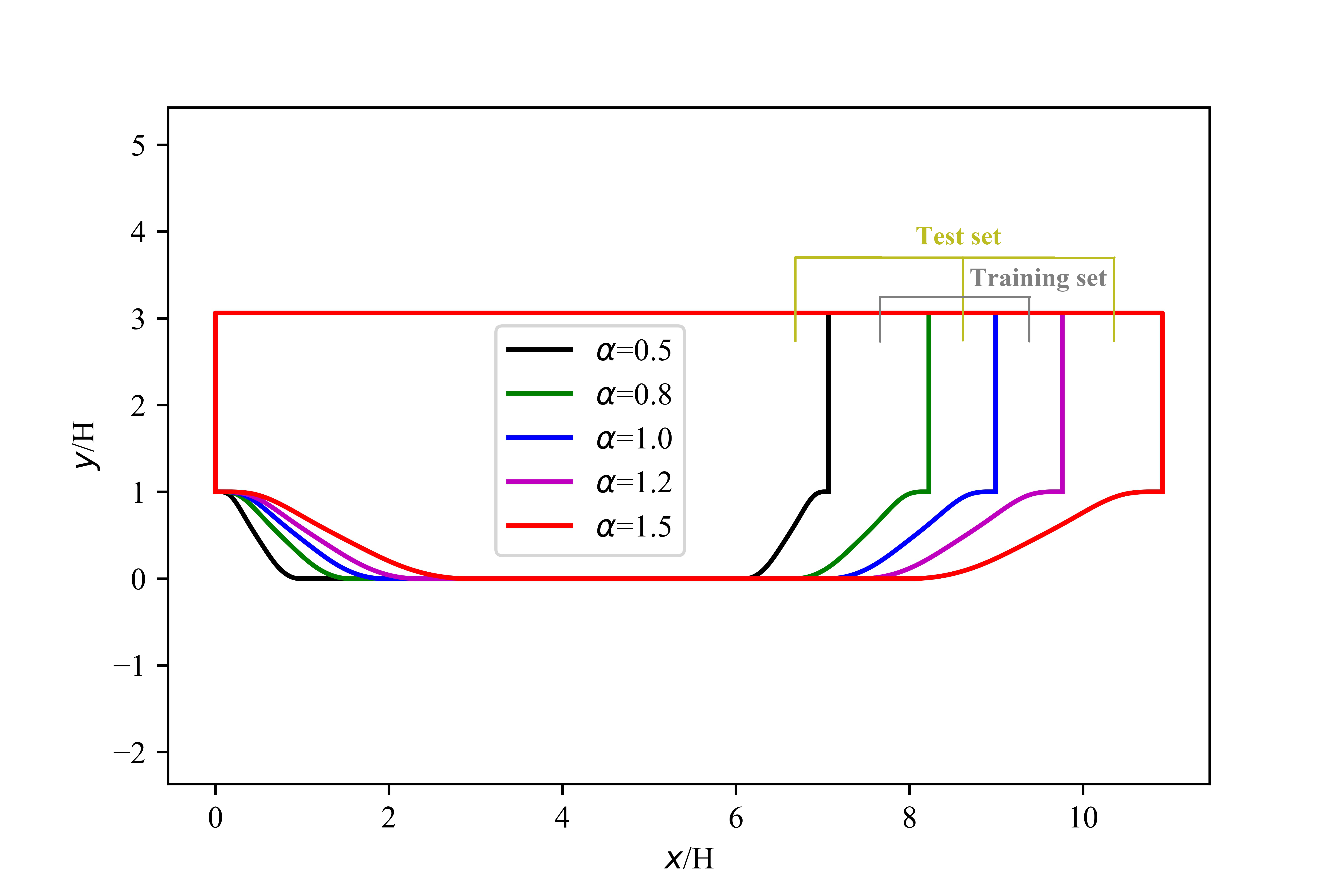} \caption{Parameterized period hill geometries with different $\alpha$.}\label{Parameterized period hill geometries with different}
\end{figure*}

In this study, we employ DNS data of parameterized periodic hill flows from \citet{xiao_flows_2020} as our training dataset. The geometries of these parameterized periodic hills with varying $\alpha$ values are illustrated in figure \ref{Parameterized period hill geometries with different}. For our analysis, we utilize data with $\alpha = 0.8$ and $\alpha = 1.2$ as the training set while reserving $\alpha = 0.5$, $1.0$, and $1.5$ for part of the test set. This experimental design allows us to evaluate our model's performance on interpolated ($\alpha = 1.0$) and extrapolated ($\alpha = 0.5$ and $1.5$) flow conditions.

It is worth noting that the training flow dataset employed in this study is not two-dimensional. The flow data from \citet{xiao_flows_2020} are 2D slices extracted from DNS results of 3D periodic hill flows. Because the spanwise velocity component, which is perpendicular to the flow domain illustrated in figure \ref{Parameterized period hill geometries with different}, can exceed the magnitude of the $y$-direction velocity component in certain regions, the training flow dataset used in this study is inherently three-dimensional.

Table \ref{tab:test_cases} presents the flow scenarios for the whole test case. These test cases enable a comprehensive assessment of the generalizability of our symbolic regression-based turbulence models.

\begin{table}
\begin{tabular}{p{3cm}p{4.5cm}p{5.5cm}}
Test case flow scenario & 
Flow characteristics & 
Description \\ [3pt]
Periodic hill & 
2D, Incompressible, Low Reynolds number, Flow separation induced by bluff body & 
Separated flow over a bluff body with geometry analogous to the training set. This case evaluates the predictive capability of the symbolic regression model for geometries closely resembling those in the training dataset. \\

Zero pressure gradient flat plate & 
2D, Incompressible, Low Reynolds number & 
Evaluate whether the symbolic regression turbulence model adversely affects the baseline model's prediction accuracy for wall-bounded flow quantities. \\

NACA 0012 airfoil incompressible flow & 
2D, Incompressible, High Reynolds number, Flow separation induced by bluff body & 
Separated flow over a bluff body featuring a geometry and Reynolds number distinctly different from those in the training set. \\

NASA Rotor 37 transonic axial compressor rotor flow & 
3D, Compressible, Transonic, Separated flow induced by bluff-body geometry and adverse pressure gradient & 
Complex engineering flows significantly different from those in the training set, representing realistic turbomachinery operating conditions. \\

\end{tabular}
\caption{Test case flow scenarios}
\label{tab:test_cases}
\end{table}

\section{Results}
\label{sec: Results}

\subsection{Symbolic regression results}
\label{sec: Symbolic regression results}

Eq.~(\ref{eq: SR_results}) presents the symbolic regression results of our turbulence model. The $g_1 = -\frac{v_t}{k}$ is not obtained through symbolic regression training, but is instead fixed a prior according to $\S$ \ref{sec: Symbolic regression}. It can be observed that the expressions obtained for $\hat{g}_3$ and $\hat{g}_4$ have opposite signs. Moreover, the tensor invariants and additional features that appear in the symbolic regression results are $I_3$, $I_4$, $I_5$, $q_2$, $q_4$, and $q_5$.

\begin{align}
\left\{\begin{array}{llll}
g_1 = -\frac{v_t}{k} \\
\hat{g}_2 = 0.143 q_2 -0.418  \\
\hat{g}_3 =  I_5 - I_3\\
\hat{g}_4 = I_3 - I_5 \\
\hat{g}_6 = (I_4 + q_4) q_5
\end{array}\right.
\label{eq: SR_results}
\end{align}

The tensor invariants $I_3$ can be formulated as:
\begin{align}
I_3=\operatorname{Tr}\left(\hat{\mathsfbi{R}}^2\right)=\operatorname{Tr}\left[\left(\frac{\mathsfbi{R}}{\|\mathsfbi{R}\|_F+\omega}\right)^2\right]=-\left(\frac{\|\mathsfbi{R}\|_F}{\|\mathsfbi{R}\|_F+\omega}\right)^2=-\left(\frac{1/\omega}{1/\omega+1/\|\mathsfbi{R}\|_F}\right)^2.
\label{eq: I3}
\end{align}
The quantity $I_3$ lies in the interval $[-1,0]$. The term $1/\omega$ represents the turbulence time scale, whereas $1/||\mathsfbi{R}||_F$ denotes the mean rotation time scale. Accordingly, this parameter characterizes the negative ratio between the turbulence time scale and the mean rotation time scale. When $1/\omega \gg 1/||\mathsfbi{R}||_F$, $I_3$ converges to $-1$, indicating that the turbulence time scale dominates over the mean rotation time scale. In contrast, when $1/||\mathsfbi{R}||_F \gg 1/\omega$, $I_3$ approaches $0$, implying that the mean rotation time scale is significantly larger than the turbulence time scale.

The $I_4$ can be expressed as:
\begin{align}
I_4 = \operatorname{Tr}\left(\hat{\mathsfbi{A}}_p^2\right) = \operatorname{Tr}\left[\left(\frac{\mathsfbi{A}_p}{\left\|\mathsfbi{A}_p\right\|_F+\omega \sqrt{k}}\right)^2\right] = -2\left(\frac{\|\nabla p\|_2}{\sqrt{2}\|\nabla p\|_2+\omega \sqrt{k}}\right)^2.
\label{eq: I4}
\end{align}
The quantity $I_4$ lies in the interval $[-1,0]$. When $||\nabla p||_2 \gg \omega \sqrt{k}$, $I_4$ approaches $-1$, indicating that the norm of the mean pressure gradient is much larger than the characteristic acceleration of turbulent fluctuations. Conversely, when $\omega \sqrt{k} \gg ||\nabla p||_2$, $I_4$ approaches $0$, indicating that the characteristic acceleration of turbulent fluctuations dominates over the norm of the mean pressure gradient.

The $I_5$ can be expressed as:
\begin{align}
I_5 = \operatorname{Tr}\left(\hat{\mathsfbi{A}}_k^2\right) = \operatorname{Tr}\left[\left(\frac{\mathsfbi{A}_k}{\left\|\mathsfbi{A}_k\right\|_F+\omega \sqrt{k}}\right)^2\right] = -2\left(\frac{\|\nabla k\|_2}{\sqrt{2}\|\nabla k\|_2+\omega \sqrt{k}}\right)^2.
\label{eq: I5}
\end{align}
The $I_5$ lies in the interval $[-1,0]$. When $||\nabla k||_2 \gg \omega \sqrt{k}$, $I_5$ approaches $-1$, indicating that the norm of the turbulent kinetic energy gradient is much larger than the characteristic acceleration of turbulent fluctuations. Conversely, when $\omega \sqrt{k} \gg ||\nabla k||_2$, $I_5$ approaches $0$, indicating that the characteristic acceleration of turbulent fluctuations dominates over the norm of the turbulent kinetic energy gradient.

For $q_2 = \min \left(\frac{\sqrt{k} d}{50 v}, 2\right)$, its value lies in the range $[0, 2]$. For each flow case considered in this study, the molecular viscosity is constant. Consequently, when $\sqrt{k}, d$ is sufficiently large, $q_2$ attains its upper limit of $2$, whereas when $\sqrt{k}d = 0$, we have $q_2 = 0$.

The $q_4$ can be expressed as:
\begin{align}
q_4=\frac{k}{k+\nu\|\mathsfbi{S}\|_F}.
\label{eq: q4}
\end{align}
It can be expressed as the ratio of the Reynolds stress intensity to the molecular viscous stress intensity, with $q_4 \in [0,1]$. When $k \gg \nu||\mathsfbi{S}||_F$, $q_4$ approaches 1, indicating that the Reynolds stress intensity is much greater than the molecular viscous stress intensity. Conversely, when $\nu||\mathsfbi{S}||_F \gg k$, $q_4$ approaches 0, indicating that the molecular viscous stress intensity dominates over the Reynolds stress intensity.

The $q_5$ can be expressed as:
\begin{align}
q_5=\frac{\|\mathsfbi{S}\|_F}{\|\mathsfbi{S}\|_F+\omega}=\frac{1/\omega}{1/\omega+1/\|\mathsfbi{S}\|_F}.
\label{eq: q5}
\end{align}
It can be shown that $q_5 \in [0,1]$. The quantity $1/||\mathsfbi{S}||_F$ represents the mean strain time scale, whereas $1/\omega$ represents the turbulence time scale. When $1/\omega \gg 1/||\mathsfbi{S}||_F$, $q_5$ approaches 1, indicating that the mean strain time scale is much smaller than the turbulence time scale. Conversely, when $1/||\mathsfbi{S}||_F \gg 1/\omega$, $q_5$ approaches 0, indicating that the turbulence time scale is much smaller than the mean strain time scale.

Based on the above analysis of tensor invariants and additional features associated with the symbolic regression results, we further interpret the symbolic regression model given in Eq.~(\ref{eq: SR_results}). For $\hat{g}_2$, it increases monotonically with $q_2$, and its values lie in the range $\hat{g}_2 \in [-0.418,-0.132]$. The functions $\hat{g}_3$ and $\hat{g}_4$ are bounded within the interval $[-1,1]$. For $\hat{g}_3$, when $I_5 = -1$ and $I_3 = 0$, indicating that the norm of the turbulent kinetic energy gradient is much larger than the characteristic acceleration of turbulent fluctuations and that the mean rotation time scale markedly exceeds the turbulence time scale, $\hat{g}_3$ attains the value $-1$. Conversely, when $I_5 = 0$ and $I_3 = -1$, implying that the characteristic acceleration of turbulent fluctuations is much larger than the norm of the turbulent kinetic energy gradient and that the turbulence time scale markedly exceeds the mean rotation time scale, $\hat{g}_3$ reaches the value $1$. Since $\hat{g}_4$ exhibits the opposite behavior to $\hat{g}_3$, its physical interpretation follows directly from that of $\hat{g}_3$. For $\hat{g}_6$, its values are also confined to the interval $[-1,1]$. When $\hat{g}_6 = -1$, corresponding to $I_4 = -1$, $q_4 = 0$, and $q_5 = 1$, the norm of the mean pressure gradient is much larger than the characteristic acceleration of turbulent fluctuations, the molecular viscosity stress intensity dominates the Reynolds stress intensity, and the mean strain time scale is much smaller than the turbulence time scale. In contrast, when $\hat{g}_6 = 1$, corresponding to $I_4 = 0$, $q_4 = 1$, and $q_5 = 1$, the characteristic acceleration of turbulent fluctuations dominates over the norm of the mean pressure gradient, the Reynolds stress intensity is much larger than the molecular viscosity stress intensity, and the mean strain time scale remains much smaller than the turbulence time scale

\subsection{Periodic hill flows}
\label{sec: Periodic hill flows}

\begin{figure}
\centering
\includegraphics[width=1.0\textwidth]{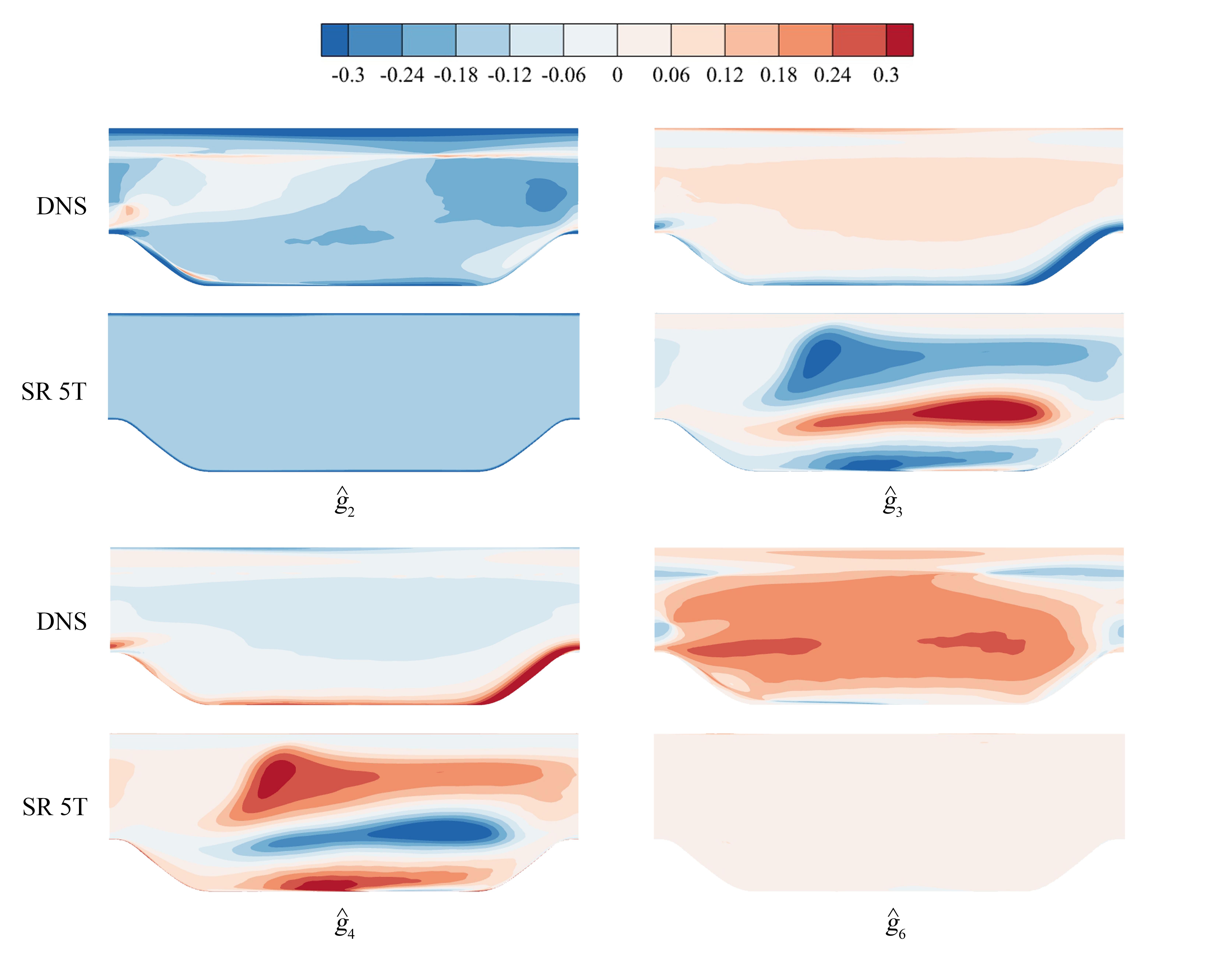}
\caption{Contour plots of the normalized tensor basis coefficients from prior results.}
\label{Contour plots of the normalized tensor basis coefficients from prior results}
\end{figure}

The computational domain for the periodic hill flow is shown in figure~\ref{Parameterized period hill geometries with different}, and the RANS mesh is adopted from \citet{mcconkey_curated_2021}. Figure \ref{Contour plots of the normalized tensor basis coefficients from prior results} presents the contour plots of the normalized tensor basis coefficients obtained from the prior results. For $\hat{g}_2$, the symbolic regression algorithm successfully captures the region of high absolute values near the wall, whereas other regions are poorly represented. For $\hat{g}_3$ and $\hat{g}_4$, the algorithm fails to achieve satisfactory predictions. For $\hat{g}_6$, the predicted values are nearly zero over most of the domain, with significant discrepancies relative to the DNS reference data. Although the prior results of the SR 5T model are not promising, the posterior results presented later in this section will demonstrate that our symbolic regression models do capture essential turbulence physics from the data.

\begin{figure}
\centering \includegraphics[width=1.0\textwidth]{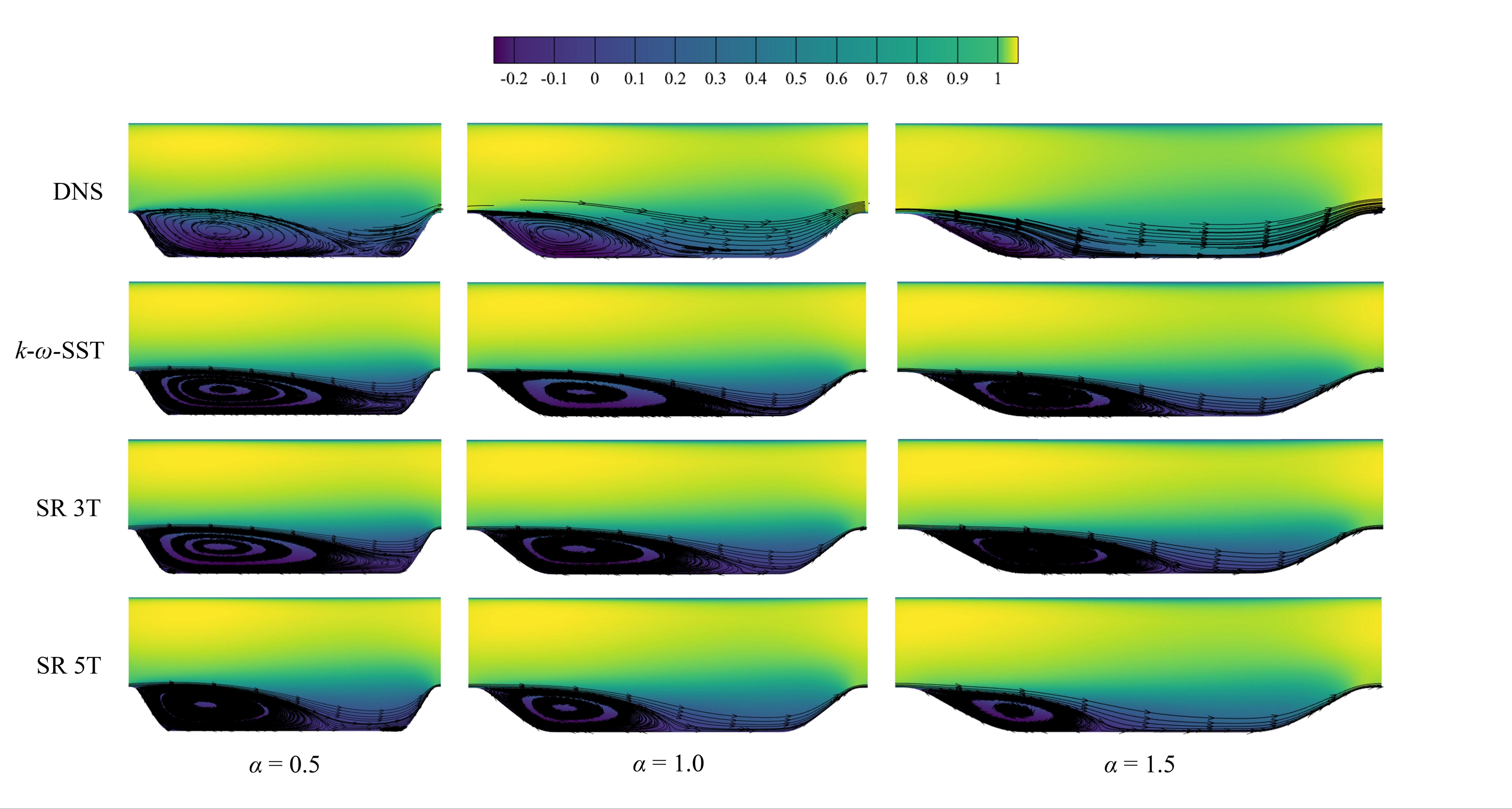} \caption{Streamlines from various simulation results of the test sets. The contours represent $U_x/U_b$.}\label{Streamline}
\end{figure}

Figure \ref{Streamline} shows the streamlines from various simulation results of the test sets. The contours represent $U_x/U_b$. It is shown that the baseline $k$-$\omega$-SST and SR 3T models yield results that are pretty similar, and both generally over-predict the separation zone in these periodic hill flows. For the SR 5T model, it underestimates the separation-zone size at $\alpha = 0.5$ and slightly overestimates it at $\alpha = 1.5$, while it predicts quite well at $\alpha = 1.0$.

\begin{figure}
\centering 
\begin{minipage}{0.7\textwidth}
  \centering
  \includegraphics[width=\textwidth]{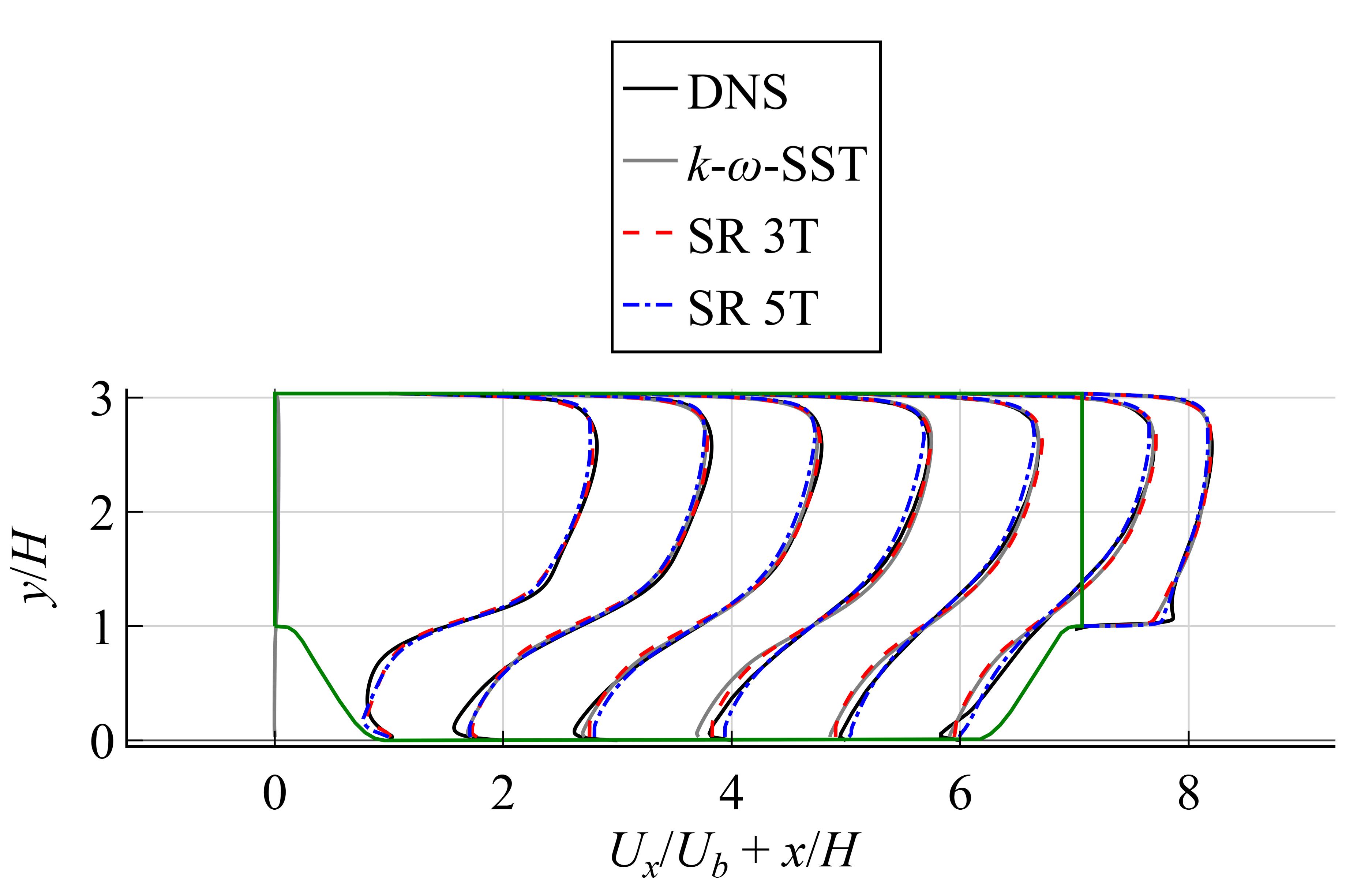}
  \centerline{(a)}
\end{minipage}
\hfill
\begin{minipage}{0.7\textwidth}
  \centering
  \includegraphics[width=\textwidth]{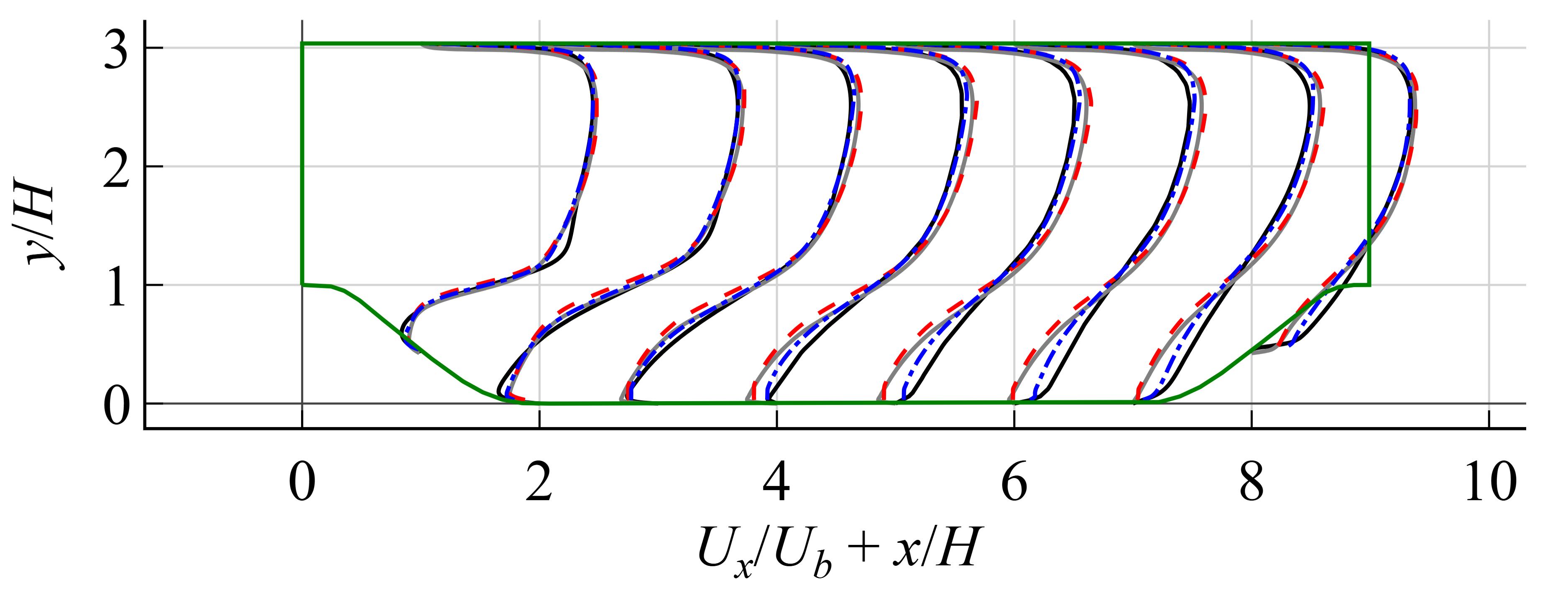}
  \centerline{(b)}
\end{minipage}

\vspace{1em}
\begin{minipage}{0.7\textwidth}
  \centering
  \includegraphics[width=\textwidth]{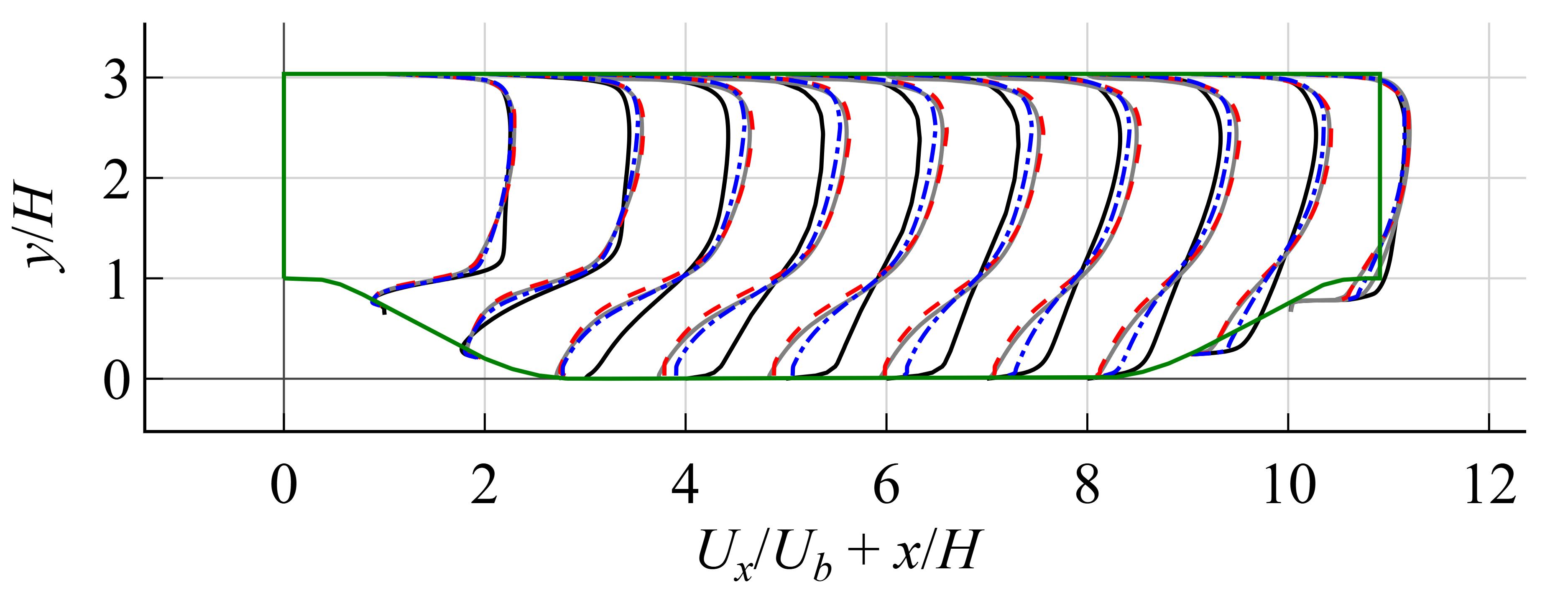}
  \centerline{(c)}
\end{minipage}
\caption{Velocity profiles in the $x$-direction across the test set from different simulation results. Panels (a), (b), and (c) present the results for the $\alpha = 0.5$, $\alpha = 1.0$, and $\alpha = 1.5$ cases, respectively.}
\label{velocity_profile_x}
\end{figure}

Figure \ref{velocity_profile_x} presents the velocity profiles in the $x$-direction across the test set obtained from different simulation approaches. Panels (a), (b), and (c) correspond to the $\alpha = 0.5$, $\alpha = 1.0$, and $\alpha = 1.5$ cases, respectively. Overall, the predictions from the baseline $k$-$\omega$-SST and SR 3T models are nearly identical. For the $\alpha = 0.5$ case, the SR 5T model yields improved predictions in the region $x/H = 4, 5, 6$ for $y/H \in (0.5, 1)$ compared to both the $k$-$\omega$-SST and SR 3T models, although it performs slightly worse in the vicinity of $y/H = 2.5$. For the $\alpha = 1.0$ case, the SR 5T model demonstrates substantially better overall performance than the $k$-$\omega$-SST and SR 3T models. Similarly, for the $\alpha = 1.5$ case, the SR 5T model generally outperforms both the $k$-$\omega$-SST and SR 3T models.

To further quantify the accuracy of our symbolic regression turbulence models, we present a comparison of the mean squared error (MSE) for the normalized velocity vector $\mathsfbi{U}/U_n$ at each grid point across various models in the test cases, where $U_n = 0.020188~\text{m/s}$ denotes the volume-averaged velocity, as summarized in Table \ref{tab: Comparison of mean squared error (MSE)}. The results indicate that the SR 3T model consistently under-performs relative to the baseline $k$-$\omega$-SST model across all cases. In contrast, the SR 5T model consistently outperforms the baseline in all cases, with particularly notable improvements observed at $\alpha = 1.0$ and $\alpha = 1.5$.

\begin{table}

\centering
\begin{tabular}{lccc}
& $k$-$\omega$-SST & SR 3T & SR 5T \\ [3pt]
$\alpha = 0.5$ & $1.57 \times 10^{-3}$ & $1.86 \times 10^{-3}$ & $1.56 \times 10^{-3}$ \\
$\alpha = 1.0$ & $5.14 \times 10^{-3}$ & $5.76 \times 10^{-3}$ & $1.60 \times 10^{-3}$ \\
$\alpha = 1.5$ & $1.28 \times 10^{-2}$ & $1.38 \times 10^{-2}$ & $4.82 \times 10^{-3}$ \\
AVE & $6.49 \times 10^{-3}$ & $7.15 \times 10^{-3}$ & $2.66 \times 10^{-3}$ \\
\end{tabular}
\caption{\label{tab: Comparison of mean squared error (MSE)} Comparison of mean squared error (MSE) for the normalized velocity vector $\mathsfbi{U}/U_n$ at each grid point across various models in the test cases, where $U_n = 0.020188~\text{m/s}$ represents the volume-averaged velocity.}
\end{table}

\begin{figure}
\centering \includegraphics[width=1.0\textwidth]{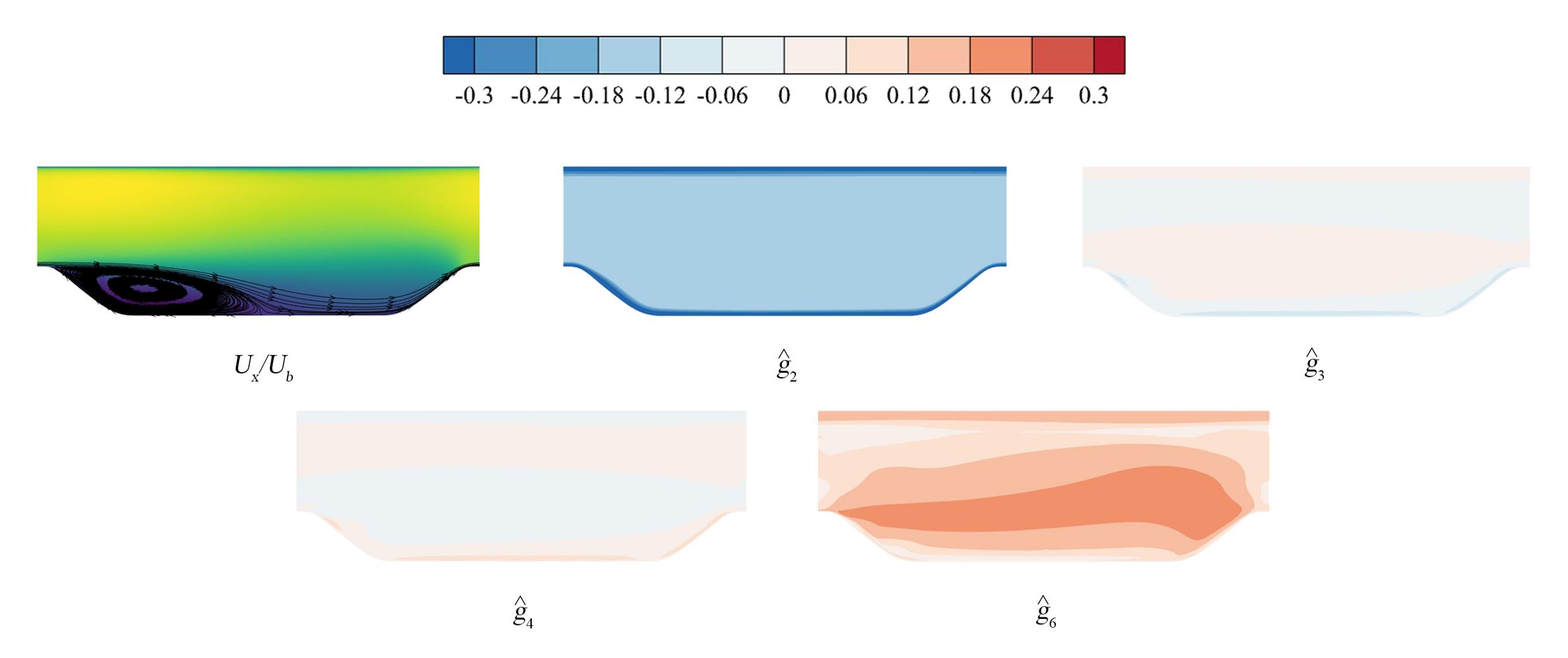} \caption{Contour plot of posterior normalized tensor basis coefficients of the SR 5T model for the $\alpha = 1.0$ case. The $\mathsfbi{U}/U_n$ contour plot is identical to that in figure \ref{Streamline}.}\label{Contour plot of normalized tensor basis coefficients}
\end{figure}

Figure \ref{Contour plot of normalized tensor basis coefficients} presents contour plots of the normalized tensor basis coefficients for the case of $\alpha = 1.0$. The $\mathsfbi{U}/U_n$ contour plot is identical to that shown in figure \ref{Streamline}. For $\hat{g}_2$, regions of large absolute values are primarily concentrated in the near-wall zone. In particular, the absolute value of $\hat{g}_2$ is higher along the upstream hill wall than in the downstream region, which can be attributed to the larger turbulence kinetic energy $k$ upstream compared with downstream. The coefficients $\hat{g}_3$ and $\hat{g}_4$ exhibit large absolute values mainly along portions of the bottom wall. For $\hat{g}_6$, regions of high absolute value are predominantly located in the strong shear region between the main stream and the separation zone. This explains why the SR 3T model does not yield an improved prediction of separation compared with the baseline $k$-$\omega$-SST model: the terms $\hat{g}_2 \mathsfbi{\hat{T}}_2$ and $\hat{g}_3 \mathsfbi{\hat{T}}_3$ are activated primarily in the near-wall zone, where they have little influence on separation. In contrast, the term $\hat{g}_6 \mathsfbi{\hat{T}}_6$ is the main contributor to the improved separation prediction of the SR 5T model, as it is predominantly activated in the strong shear region between the main stream and the separation zone.

\begin{figure}
\centering \includegraphics[width=1.0\textwidth]{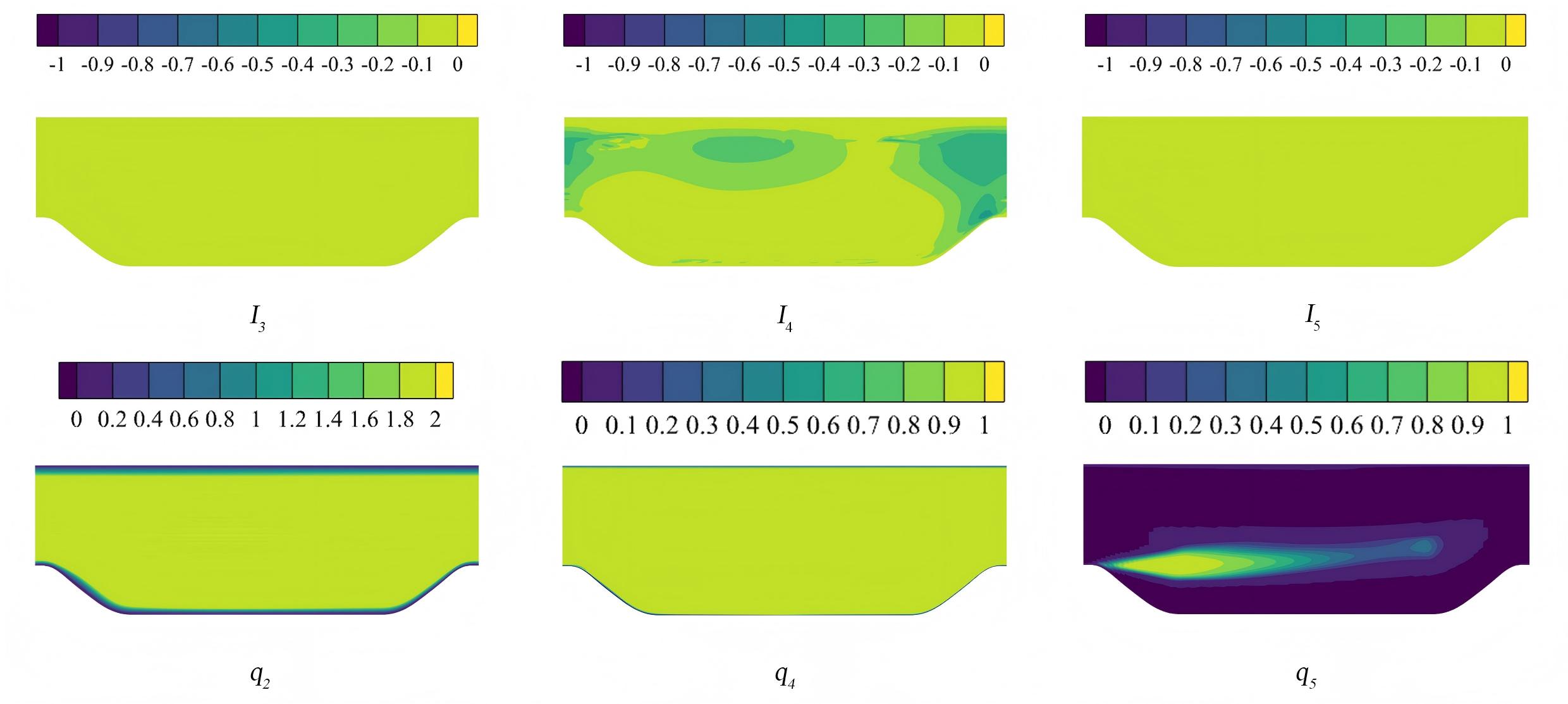} \caption{Contour plot of posterior features of the SR 5T model for the $\alpha = 1.0$ case.}\label{Contour plot of posterior features}
\end{figure}

Figure \ref{Contour plot of posterior features} presents the contour plots of the posterior features of the SR 5T model for the $\alpha = 1.0$ case. It can be observed that $I_3$ and $I_5$ are nearly zero throughout the entire flow field. For $I_4$, the values are close to zero in most regions, with some negative values appearing predominantly in the upper region of the flow. Regarding $q_2$, the values are nearly zero in the near-wall region, while they increase to approximately 2 in regions farther away from the wall. For $q_4$, the values are close to 1 over most of the domain, except in the near-wall region. In the case of $q_5$, high values are mainly concentrated in the strong shear region located between the main stream and the separation zone. Based on these observations and the results shown in Figure \ref{Contour plot of normalized tensor basis coefficients}, it can be concluded that the primary deviation from the classical Boussinesq hypothesis occurs in regions characterized by a low wall-distance-based Reynolds number and a strong mean strain rate. When the wall-distance-based Reynolds number is very large (which corresponds to regions far from the wall in most cases) and the turbulence time scale is much smaller than the mean strain time scale, the model nearly recovers the classical Boussinesq hypothesis. Furthermore, regions with large absolute values of $I_4$ and $q_5$ exhibit minimal spatial overlap.

\begin{figure}
\centering 
\begin{minipage}{0.6\textwidth}
  \centering
  \includegraphics[width=\textwidth]{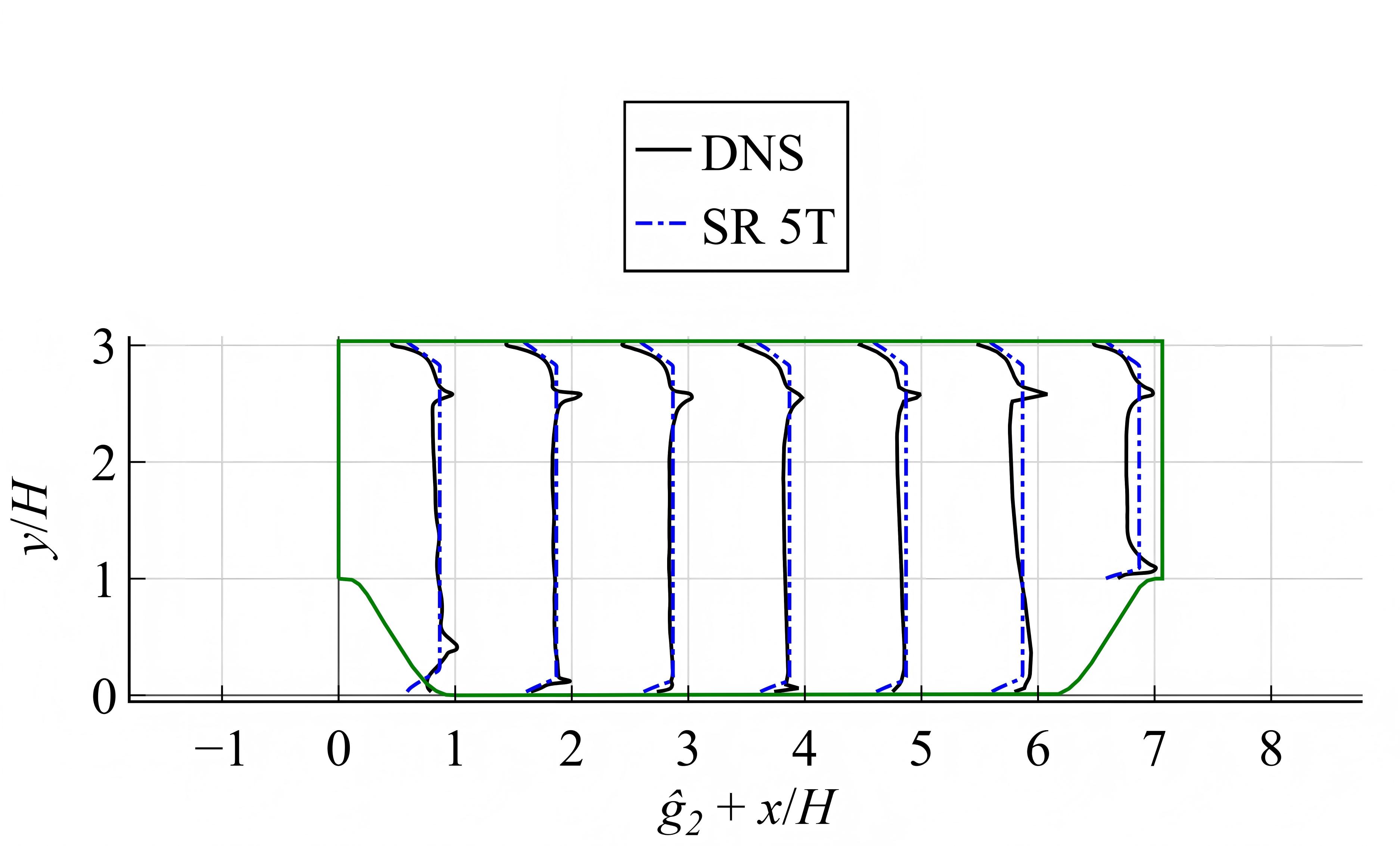}
  \centerline{(a)}
\end{minipage}
\hfill
\begin{minipage}{0.6\textwidth}
  \centering
  \includegraphics[width=\textwidth]{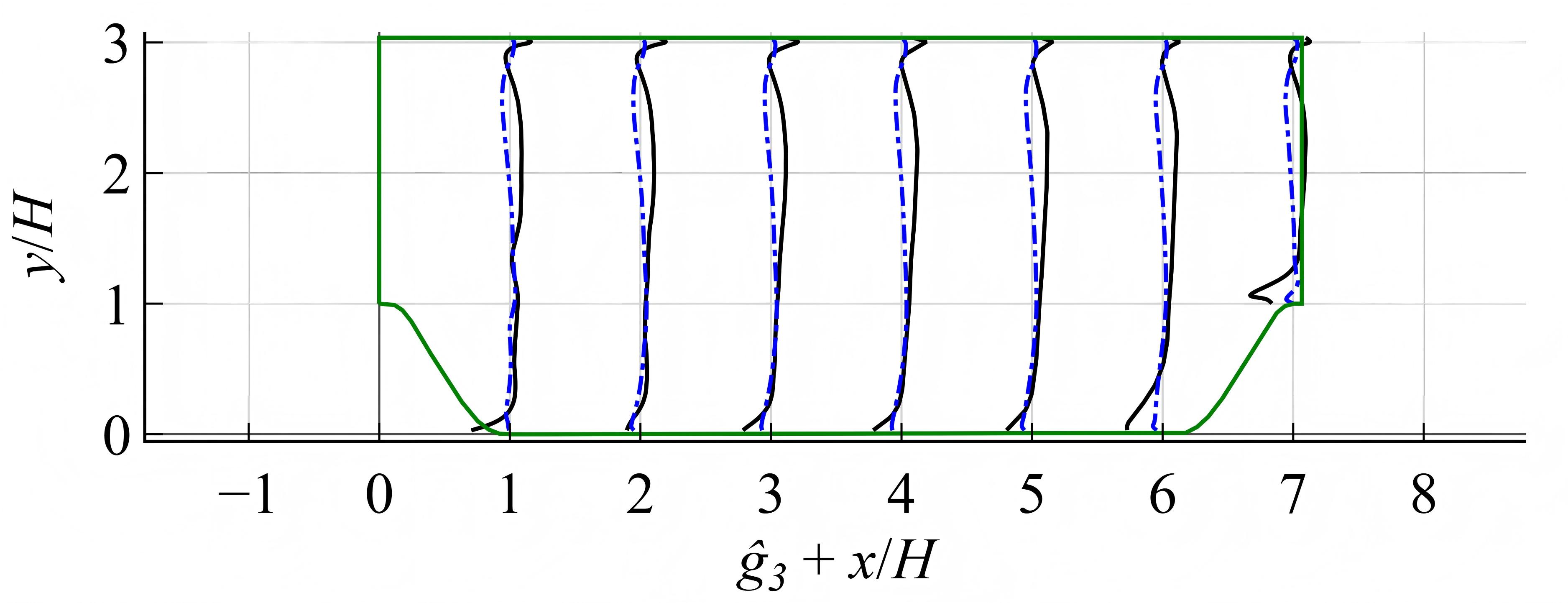}
  \centerline{(b)}
\end{minipage}

\vspace{1em}
\begin{minipage}{0.6\textwidth}
  \centering
  \includegraphics[width=\textwidth]{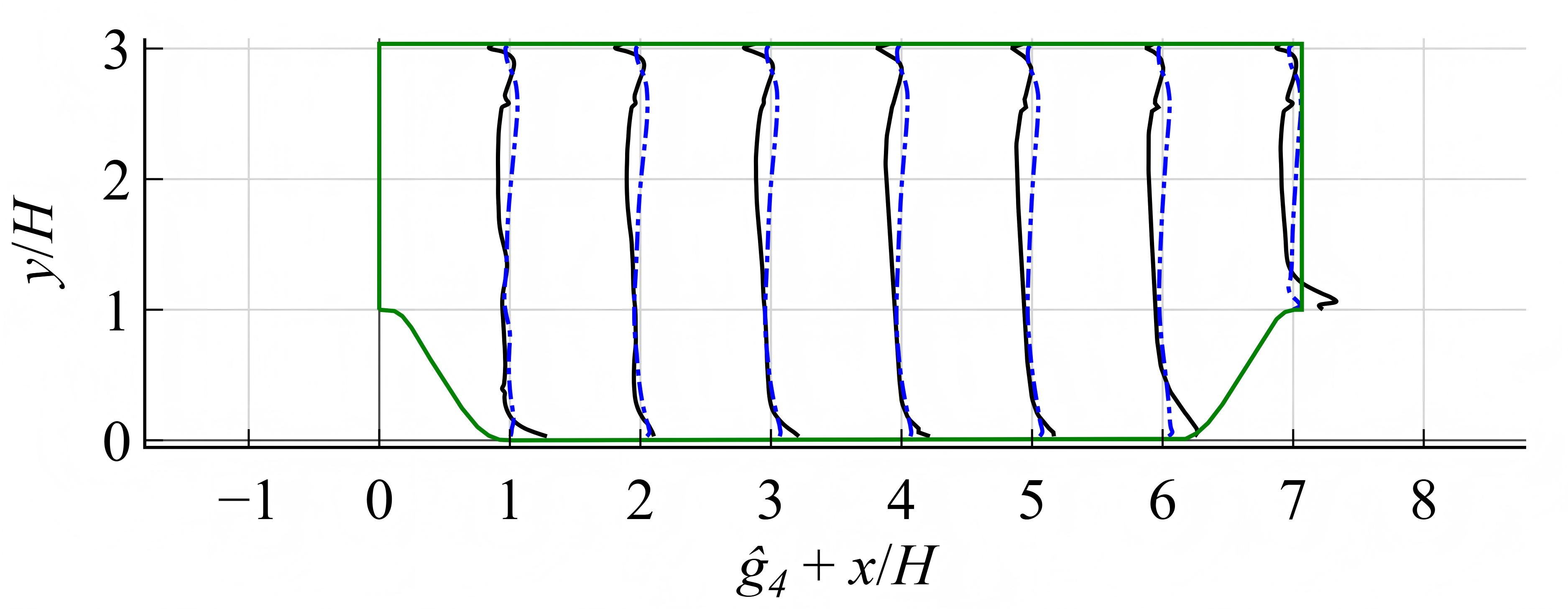}
  \centerline{(c)}
\end{minipage}

\vspace{1em}
\begin{minipage}{0.6\textwidth}
  \centering
  \includegraphics[width=\textwidth]{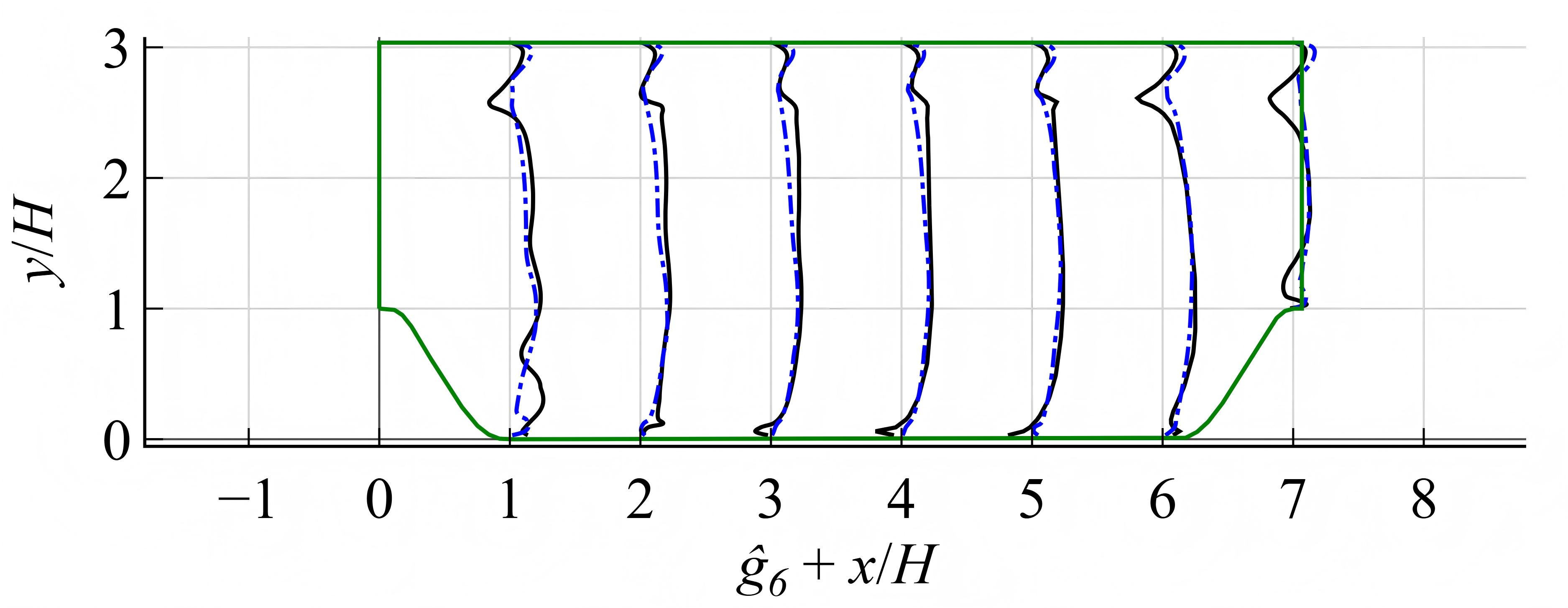}
  \centerline{(d)}
\end{minipage}
\caption{Posterior normalized tensor coefficient profiles of $\hat{g}_2$, $\hat{g}_3$, $\hat{g}_4$, and $\hat{g}_6$ of SR 5T model along the $x$-direction for the $\alpha = 0.5$ case.}
\label{PH_ghat}
\end{figure}

Figure \ref{PH_ghat} shows the posterior normalized tensor coefficient profiles of $\hat{g}_2$, $\hat{g}_3$, $\hat{g}_4$, and $\hat{g}_6$ along the $x$-direction. For $\hat{g}_2$, the SR 5T model provides good predictions in the region near $y/H = 1$, but fails to capture the bump accurately observed around $y/H = 2.5$. For $\hat{g}_3$ and $\hat{g}_4$, the SR 5T model performs well in the range $y/H \in (0.5, 1)$, whereas its predictive accuracy deteriorates in the vicinity of $y/H = 2.0$. For $\hat{g}_6$, the SR 5T model predicts the results quite well in the interval $y/H \in (1, 2)$. Overall, although the prior results shown in figure \ref{Contour plots of the normalized tensor basis coefficients from prior results} suggest that the prior model performance is not very promising, the posterior results presented in figure \ref{PH_ghat} indicate that the SR 5T model has learned essential features from the DNS data.

\subsection{Zero pressure gradient flat plate flow}
\label{sec: Zero pressure gradient flat plate flow}

\begin{figure*}
\centering \includegraphics[width=0.8\textwidth]{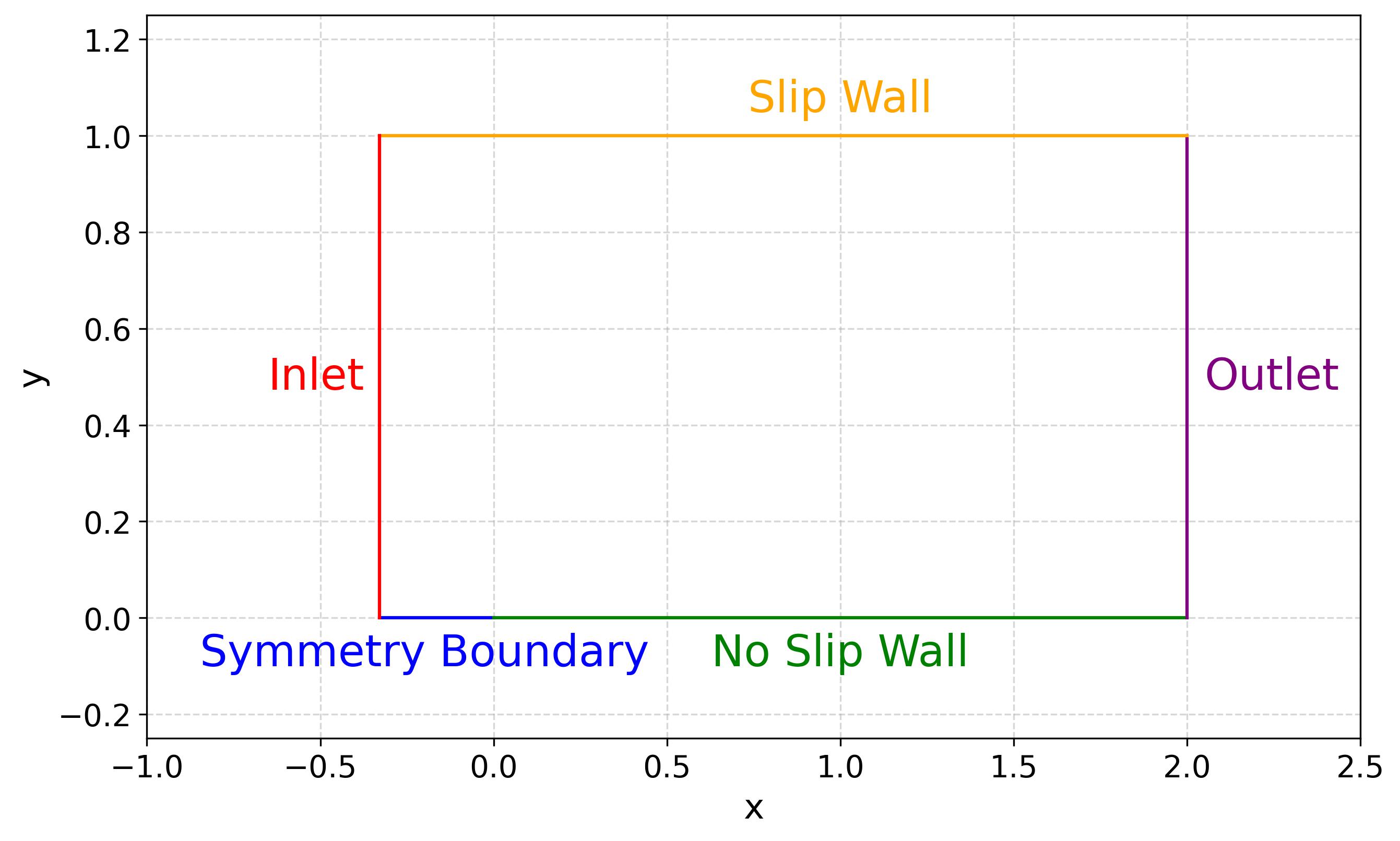} \caption{Computational domain for the zero pressure gradient flat plate flow simulation.}\label{flat_plate}
\end{figure*}

The zero pressure gradient flat plate flow is characterized by a Reynolds number, based on the plate length, of $Re = 5 \times 10^6$. The computational domain is illustrated in Fig. \ref{flat_plate}. The mesh configuration is adopted from the Turbulence Modeling Resource, with a resolution of $545 \times 385$ (including 449 points along the solid plate). The near-wall resolution corresponds to $y^+ \sim 0.1$, and the mesh has been validated by \citet{jespersen2016turbulence}.

\begin{figure}
\centering
\begin{minipage}{0.48\textwidth}
  \centering
  \includegraphics[width=\textwidth]{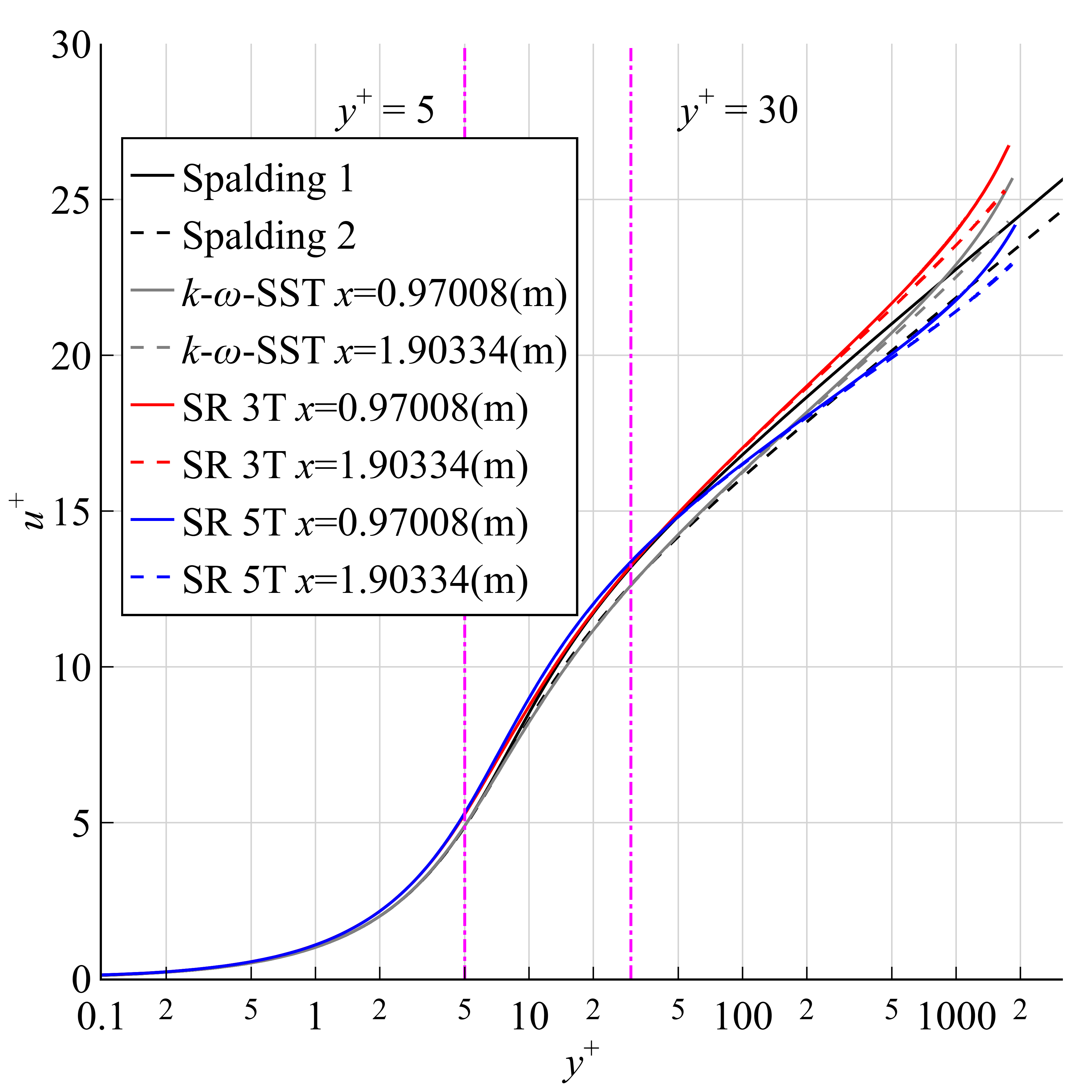}
  \centerline{(a)}
\end{minipage}
\hfill
\begin{minipage}{0.48\textwidth}
  \centering
  \includegraphics[width=\textwidth]{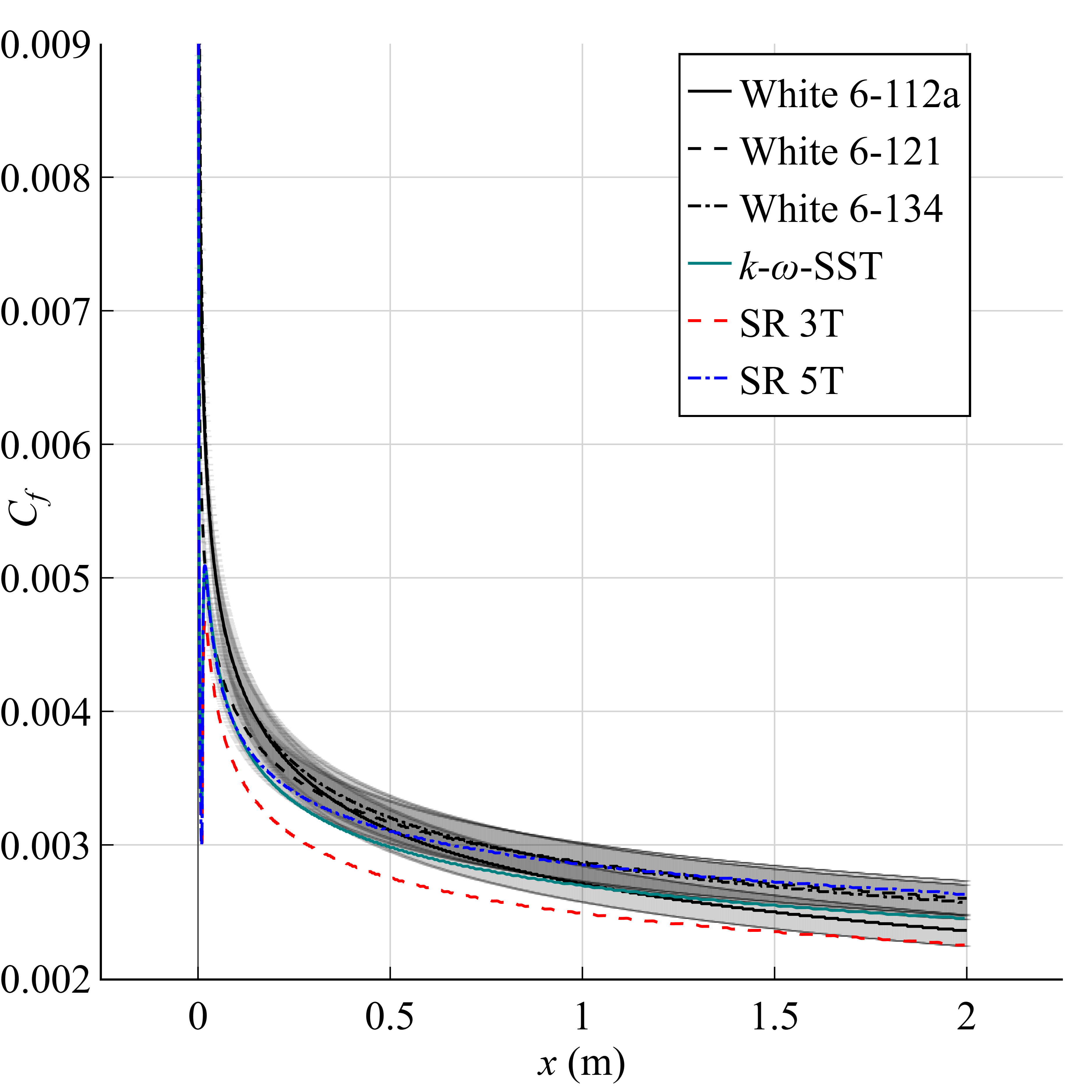}
  \centerline{(b)}
\end{minipage}
\caption{Comparative analysis of (a) law of the wall results and (b) the friction coefficient $C_f$. The gray band in (b) denotes the 5\% uncertainty envelope associated with these three theoretical results.}
\label{law_and_cf}
\end{figure}

Figure \ref{law_and_cf} (a) presents a comparative analysis of the law of the wall results. The predictions from both the baseline $k$-$\omega$-SST model and the SR 5T model lie predominantly between the Spalding 1 and Spalding 2 profiles. In contrast, the SR 3T model yields values that exceed both Spalding 1 and Spalding 2 in the regime where $y^+ > 100$.

Figure \ref{law_and_cf} (b) presents a comparative analysis of the skin friction coefficient $C_f$. The gray-shaded region indicates the 5\% uncertainty envelope for the three theoretical correlations. It can be observed that the predictions from both the baseline $k$-$\omega$-SST model and the SR 5T model fall within at least one of the theoretical uncertainty envelopes. In contrast, the predictions obtained from the SR 3T model lie predominantly outside these uncertainty bounds.

\begin{figure}
\centering 
\begin{minipage}{0.48\textwidth}
  \centering
  \includegraphics[width=\textwidth]{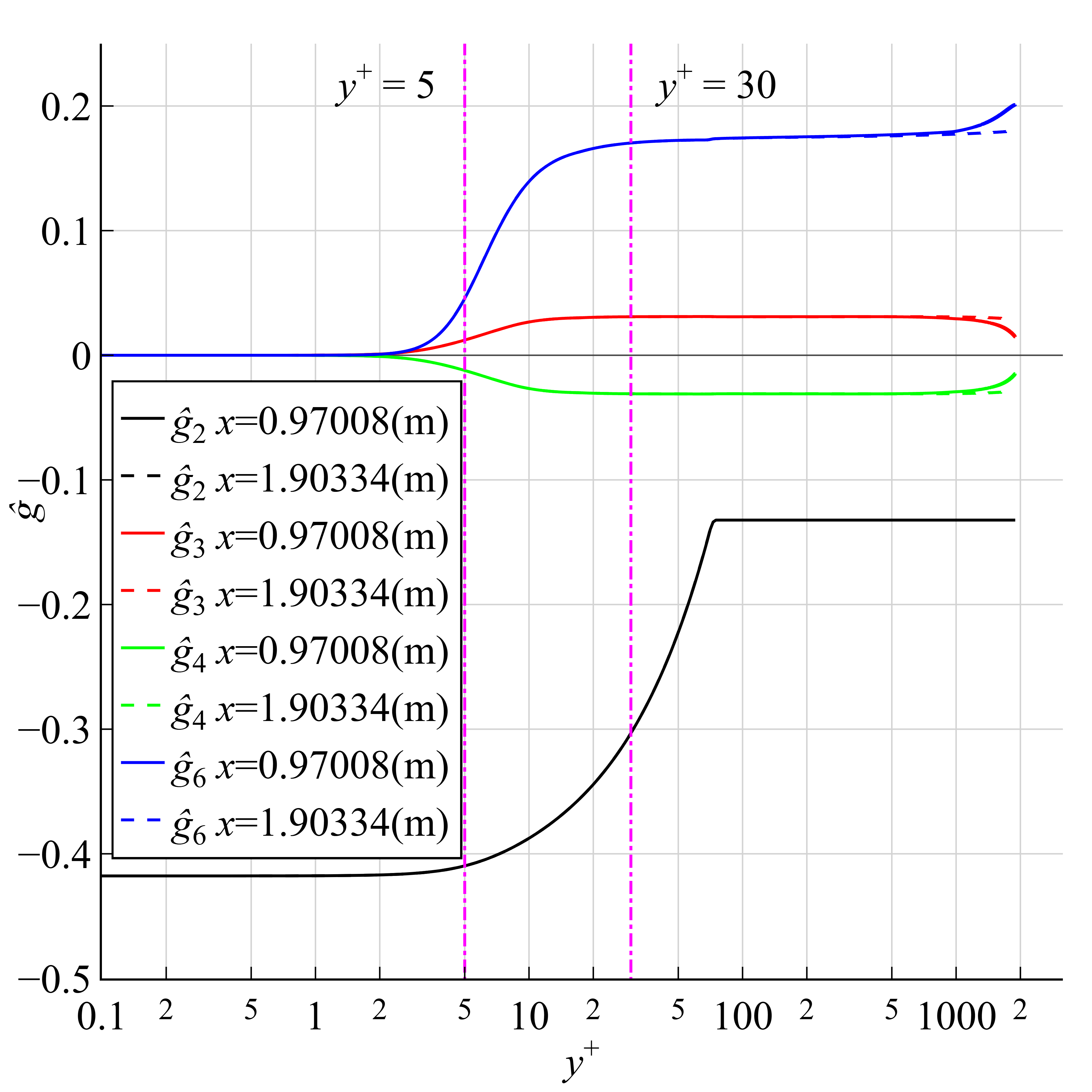}
  \centerline{(a)}
\end{minipage}
\hfill
\begin{minipage}{0.48\textwidth}
  \centering
  \includegraphics[width=\textwidth]{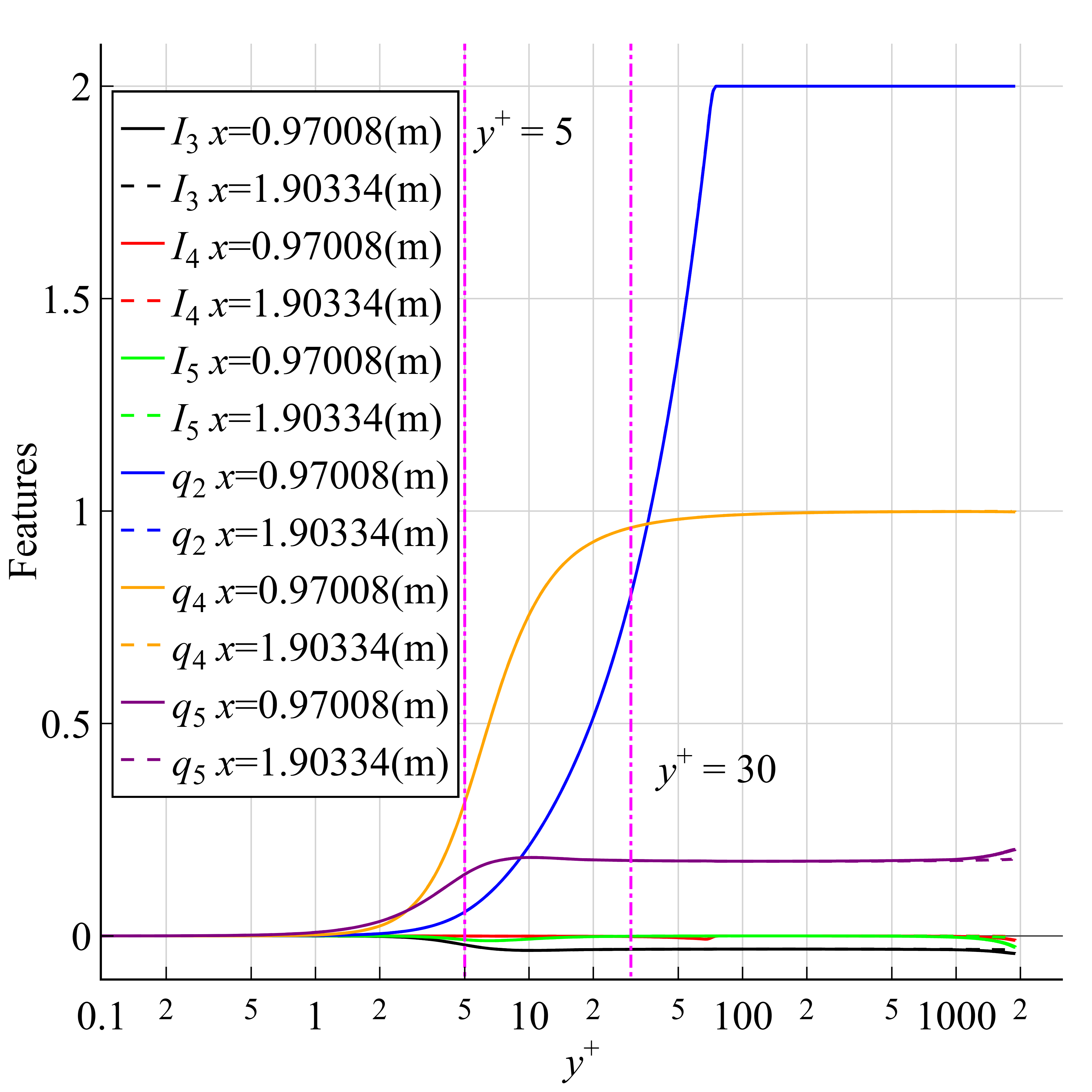}
  \centerline{(b)}
\end{minipage}
\caption{Near-wall behavior of (a) the normalized tensor basis coefficients and (b) the corresponding features.}
\label{Near-wall behavior}
\end{figure}

Figure \ref{Near-wall behavior} (a) illustrates the near-wall behavior of the normalized tensor basis coefficients. These coefficients remain approximately constant for $y^+ < 2$. In this regime, $\hat{g}_3$, $\hat{g}_4$, and $\hat{g}_6$ are nearly zero, indicating that the terms $\hat{g}_3 \mathsfbi{\hat{T}}_3$, $\hat{g}_4 \mathsfbi{\hat{T}}_4$, and $\hat{g}_6 \mathsfbi{\hat{T}}_6$ are inactive. Conversely, $\hat{g}_2$ attains its maximum absolute value in this region, suggesting that the term $\hat{g}_2 \mathsfbi{\hat{T}}_2$ plays a dominant role. For $y^+ \in (2, 1000)$, the absolute values of $\hat{g}_3$, $\hat{g}_4$, and $\hat{g}_6$ increase and eventually reach a plateau, indicating that their associated terms $\hat{g}_3 \mathsfbi{\hat{T}}_3$, $\hat{g}_4 \mathsfbi{\hat{T}}_4$, and $\hat{g}_6 \mathsfbi{\hat{T}}_6$ become relatively active in this regime. Meanwhile, the absolute value of $\hat{g}_2$ decreases and plateaus, indicating that the term $\hat{g}_2 \mathsfbi{\hat{T}}_2$ becomes relatively inactive. Notably, the onset of variation in these normalized tensor basis coefficients within the viscous sublayer occurs at approximately $y^+ = 2$, which coincides with the location where the symbolic regression models begin to deviate from the baseline $k$-$\omega$-SST model, as shown in figure \ref{law_and_cf} (a). These deviations become more pronounced in the buffer layer and log-law regions as $y^+$ increases. Furthermore, the law-of-the-wall results remain invariant along the streamwise direction, implying that the normalized tensor basis coefficients must be consistent throughout the boundary layer. Indeed, the results confirm that these coefficients are uniform within the boundary layer, and the apparent differences at different streamwise locations arise solely from variations in the local boundary-layer thickness. 

Figure \ref{Near-wall behavior} (b) illustrates the near-wall behavior of the corresponding features. These features remain nearly zero when $y^+ < 1$. For $I_3$, it decreases from zero with increasing $y^+$ and reaches a plateau for $y^+ > 10$. When $y^+ < 1.2$, $I_3$ is nearly zero, indicating that $1/||\mathsfbi{R}||_F \gg 1/\omega$. For $I_4$ and $I_5$, they are nearly zero at $x = 0.97008\,\mathrm{m}$ and $x = 1.90334\,\mathrm{m}$. For $q_2$, it increases from 0 to 2 as $y^+$ increases and reaches a plateau around $y^+ = 100$. For $q_4$, it increases from 0 to 1 with increasing $y^+$; it is nearly zero when $y^+ < 1$ and approaches unity when $y^+ > 200$, which implies that in these regimes $\nu ||\mathsfbi{S}||_F \gg k$ and $k \gg \nu ||\mathsfbi{S}||_F$, respectively. For $q_5$, it increases from zero with increasing $y^+$ and reaches a plateau when $y^+ > 10$. When $y^+ < 1$, $q_5$ is nearly zero, indicating that $1/||\mathsfbi{S}||_F \gg 1/\omega$. Based on these results and the analysis presented in $\S$ \ref{sec: Symbolic regression results}, it can be concluded that, for the SR 5T model applied to the simulation of zero pressure gradient flat plate flow, the terms $\hat{g}_3 \mathsfbi{\hat{T}}_3$ and $\hat{g}_4 \mathsfbi{\hat{T}}_4$ are inactivated when the mean rotation time scale is significantly larger than the turbulence time scale. The term $\hat{g}_6 \mathsfbi{\hat{T}}_6$ is inactivated when the molecular viscous stress intensity dominates over the Reynolds stress intensity and the turbulence time scale is much smaller than the mean strain time scale. Otherwise, these terms are activated. The term $\hat{g}_2 \mathsfbi{\hat{T}}_2$ is always activated. Therefore, when the mean rotation time scale is significantly larger than the turbulence time scale and molecular viscous stress intensity dominates over the Reynolds stress intensity (e.g., viscous sublayer), the SR 5T model nearly revert to the classical Boussinesq hypothesis.

\subsection{NACA0012 airfoil flows}
\label{sec: NACA0012 airfoil flows}

In this section, we evaluate the performance of the proposed symbolic regression-based turbulence models for incompressible flow around an NACA0012 airfoil. The simulations are performed at $Re = 3 \times 10^6$. The computational mesh is obtained from the NASA Turbulence Modeling Resource. The mesh resolution is $897 \times 257$, and detailed descriptions as well as mesh verification can be found in \citet{jespersen2016turbulence}.

\begin{figure}
\centering 
\begin{minipage}{0.3\textwidth}
  \centering
  \includegraphics[width=\textwidth]{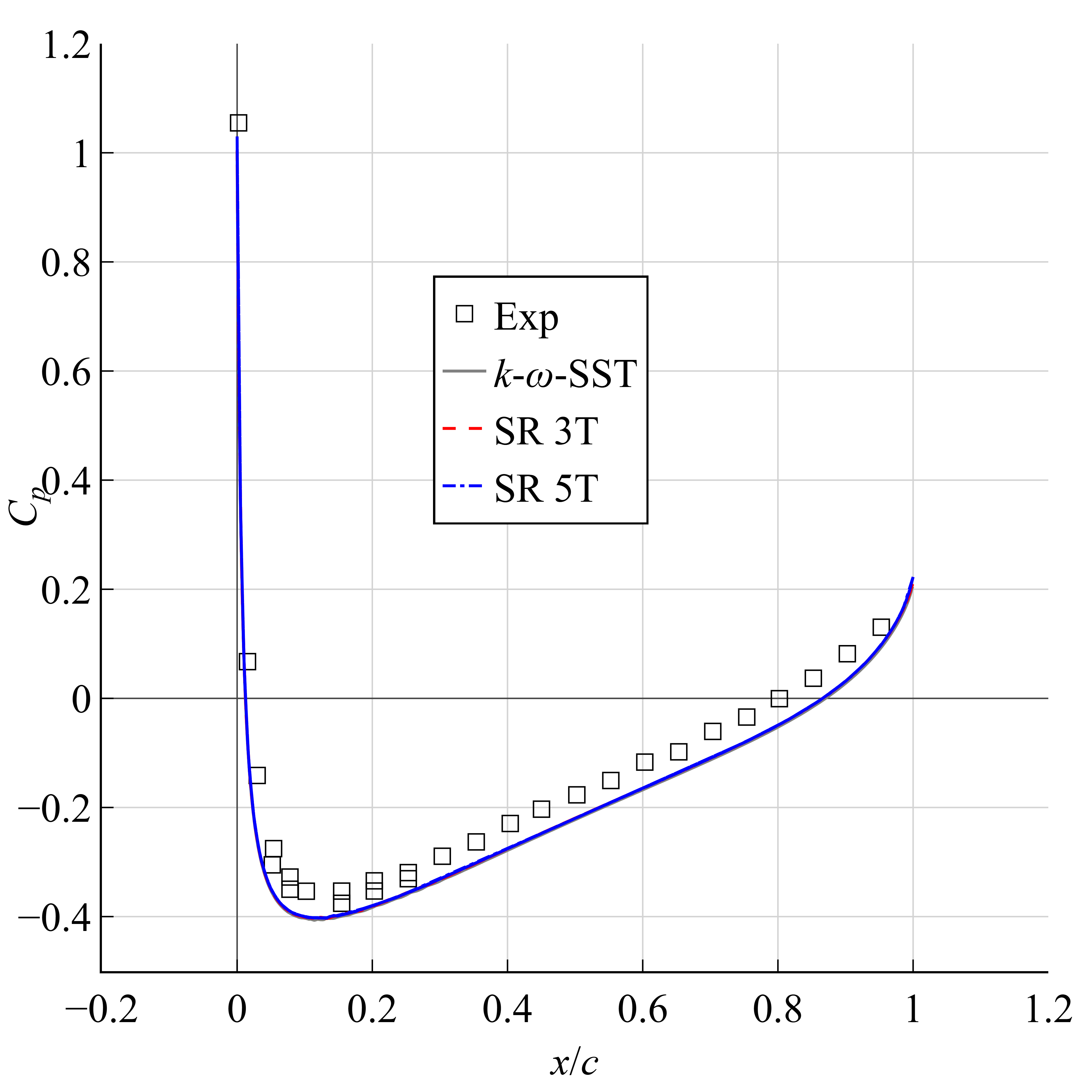}
  \centerline{(a)}
\end{minipage}
\begin{minipage}{0.3\textwidth}
  \centering
  \includegraphics[width=\textwidth]{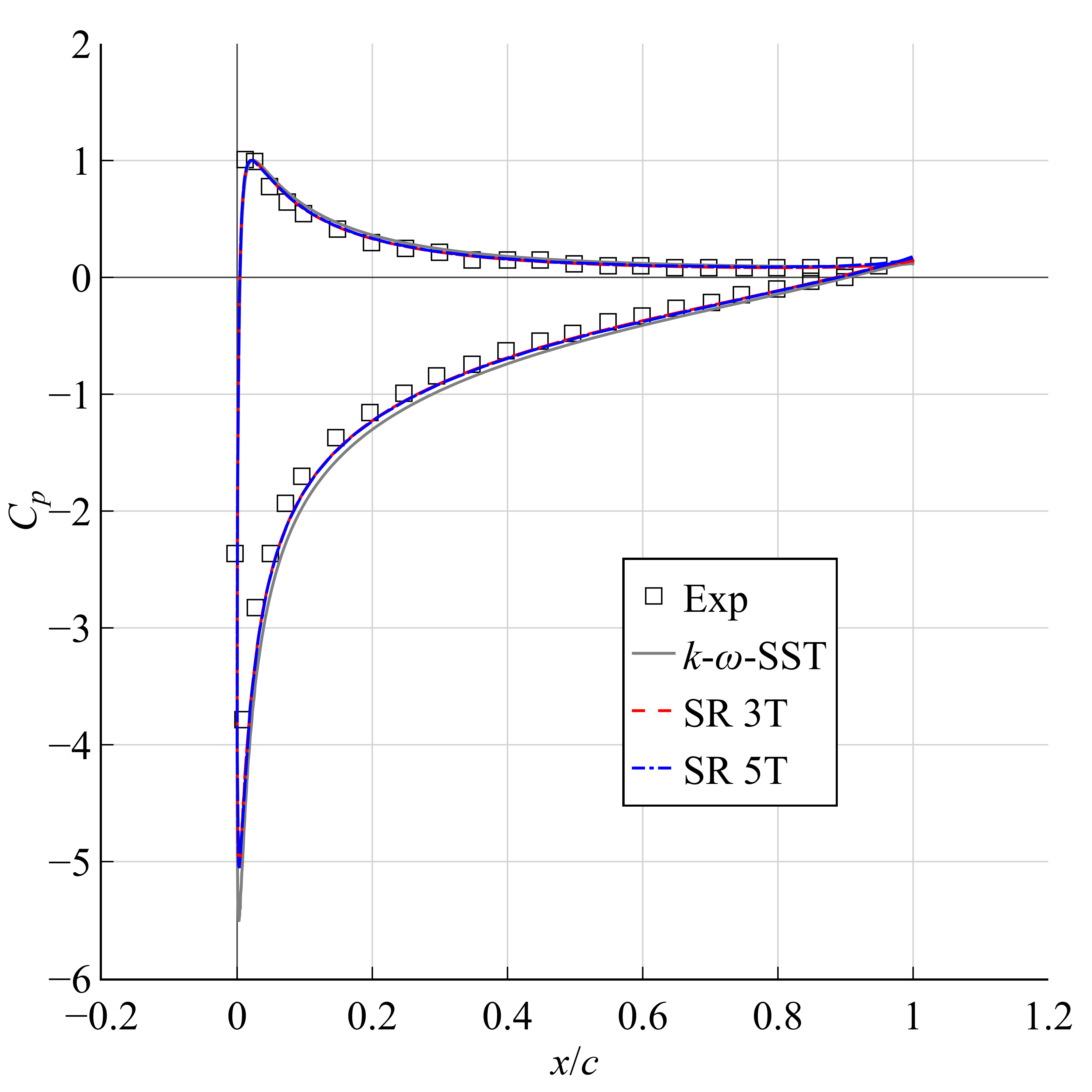}
  \centerline{(b)}
\end{minipage}
\begin{minipage}{0.3\textwidth}
  \centering
  \includegraphics[width=\textwidth]{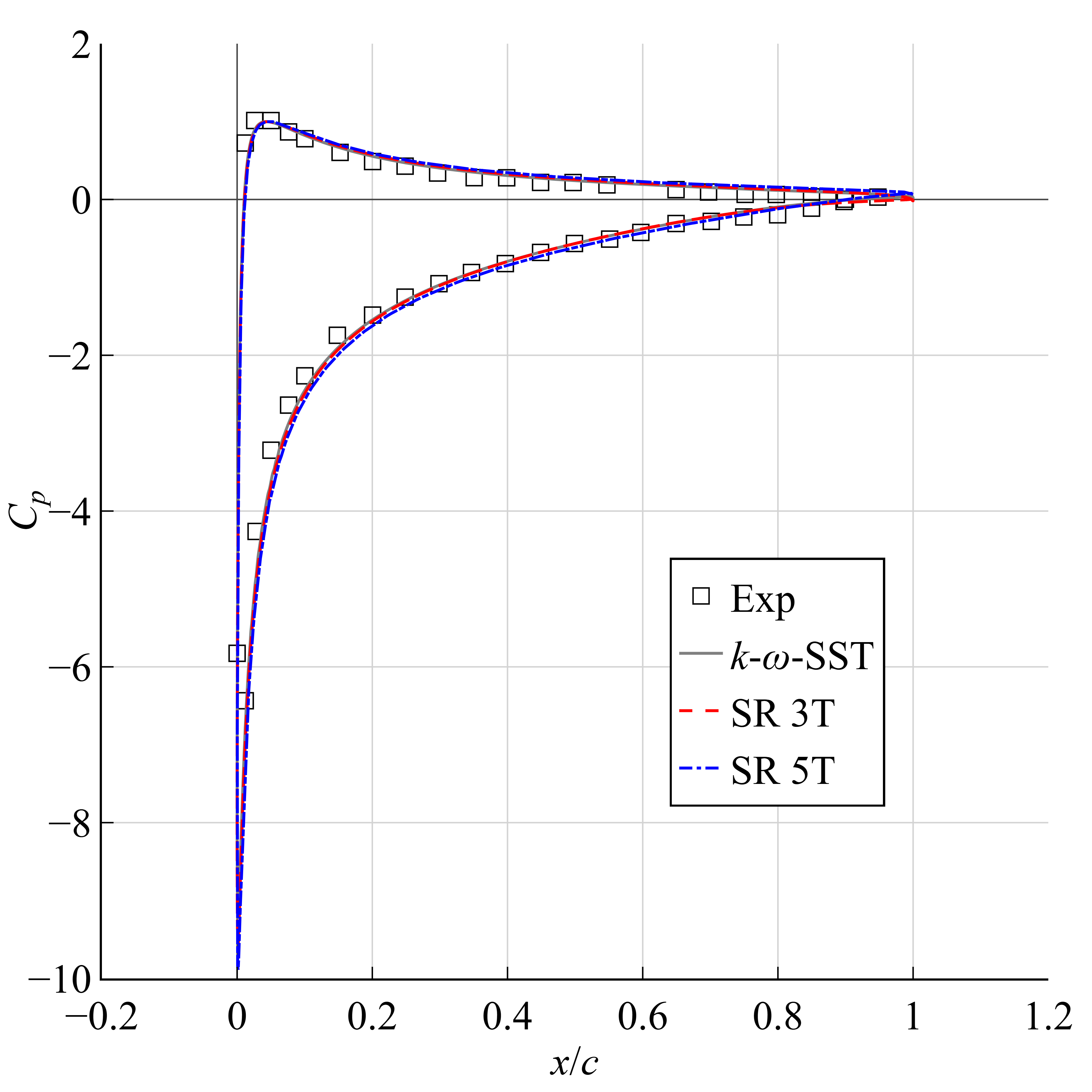}
  \centerline{(c)}
\end{minipage}
\caption{A comparative analysis of pressure coefficient distributions along the NACA0012 airfoil surface at various angles of attack (AOA). Panels (a), (b), and (c) illustrate the pressure distributions at incidence angles of $0^{\circ}$, $10^{\circ}$, and $15^{\circ}$, respectively.}
\label{NACA0012}
\end{figure}

Figure \ref{NACA0012} presents a comparative analysis of pressure coefficient distributions along the NACA0012 airfoil surface at various angles of attack (AOA). Panels (a), (b), and (c) illustrate the pressure distributions at incidence angles of $0^{\circ}$, $10^{\circ}$, and $15^{\circ}$, respectively. In these three cases, where the AOA remains relatively low, the baseline $k$-$\omega$-SST model already yields sufficiently accurate results, and our symbolic regression models maintain comparable accuracy.

\begin{figure}
\centering \includegraphics[width=0.5\textwidth]{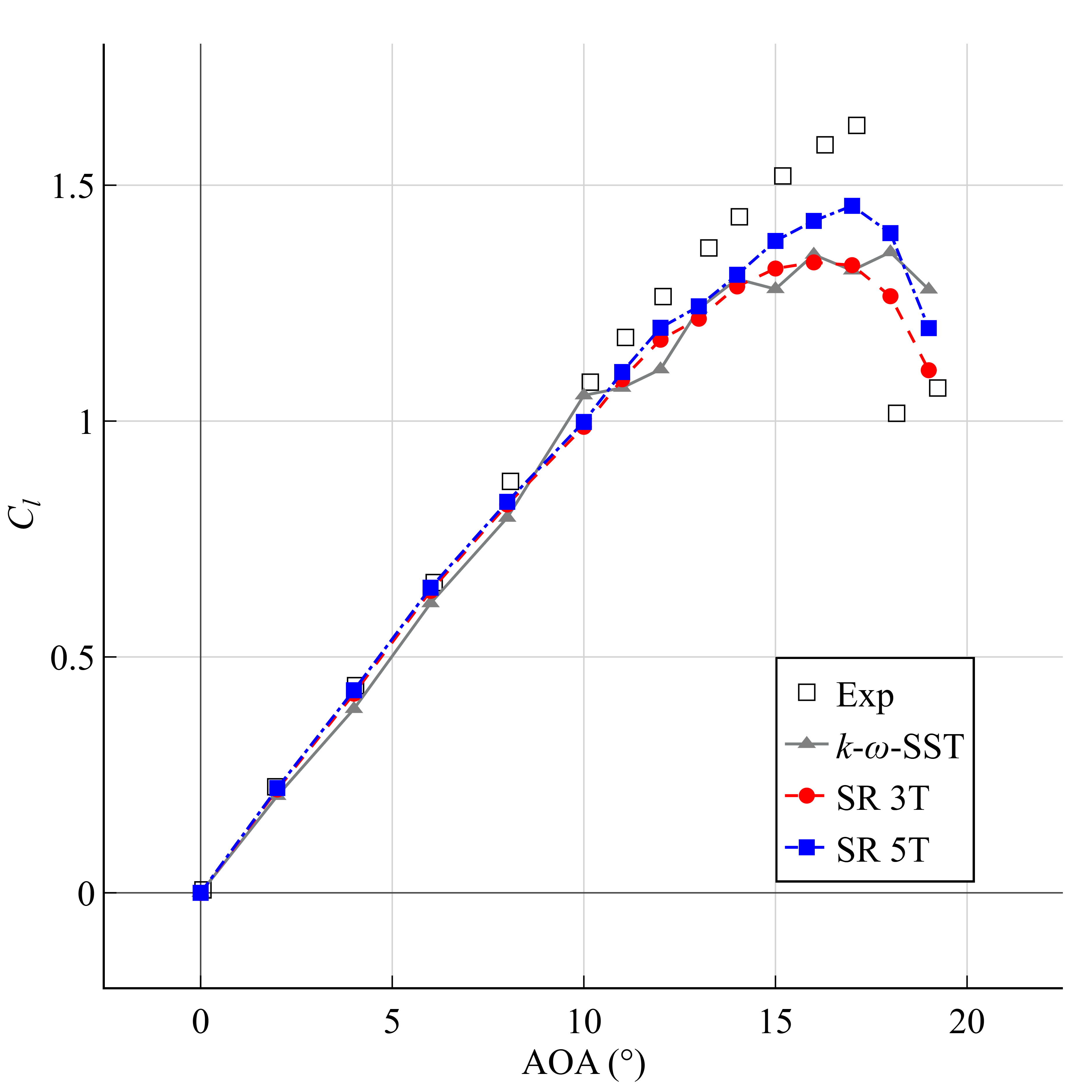} \caption{Comparative analysis of lift curve results.}\label{Comparative analysis of lift curve results}
\end{figure}

Figure \ref{Comparative analysis of lift curve results} presents a comparative analysis of the lift curve results. The results indicate that at angles of attack (AOA) of 4°, 6°, and 8°, the symbolic regression models achieve higher accuracy than the baseline $k$-$\omega$-SST model. However, at an AOA of 10°, the baseline $k$-$\omega$-SST model outperforms the symbolic regression models. For angles of attack above 11° and before the stall angle, the SR 5T model shows superior accuracy in predicting the lift coefficient compared to the other two models. In contrast, the baseline $k$-$\omega$-SST model exhibits fluctuating behavior in this regime. In the post-stall region, the SR 3T model yields better agreement with the experimental data than the other two models. Overall, the SR 5T model provides the most accurate predictions of the lift curve across the range of conditions examined.

\subsection{NASA Rotor 37 transonic axial compressor flows}
\label{sec: NASA Rotor 37 transonic axial compressor flows}

\begin{figure*}
\centering \includegraphics[width=0.8\textwidth]{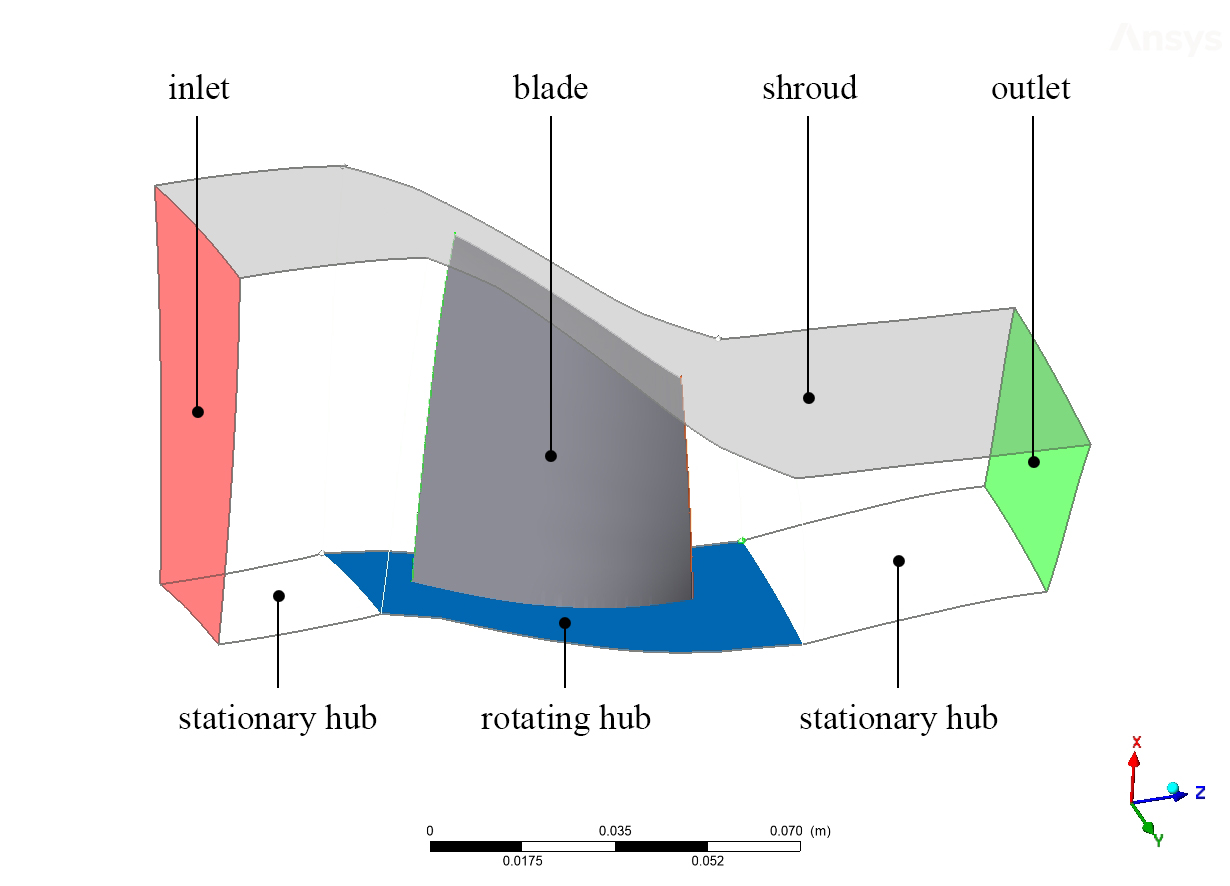} \caption{Three-dimensional computational domain of a single blade of NASA Rotor 37.}\label{Rotor_37}
\end{figure*}

In this section, we evaluate the proposed symbolic regression model using NASA Rotor 37, a transonic axial compressor rotor, as the test case. This compressor rotor leverages shock-wave mechanisms to achieve fluid compression and operates at a high rotational speed of 17188.7 rpm. It exhibits a high total pressure ratio, thereby constituting a complex, engineering-relevant turbulent flow system that differs substantially from the training data. Figure~\ref{Rotor_37} illustrates the three-dimensional computational domain corresponding to a single blade passage of NASA Rotor 37. Experimental data for this rotor were obtained from \cite{suder_experimental_1996}. The computational mesh comprises approximately 1.1 million cells, with a near-wall resolution of $y^+ \sim 1$. Detailed descriptions of the mesh generation procedure and the mesh-independence study are provided in our previous work \citep{ji_tensor_2024}.

\begin{figure}
\centering 
\begin{minipage}{0.48\textwidth}
  \centering
  \includegraphics[width=\textwidth]{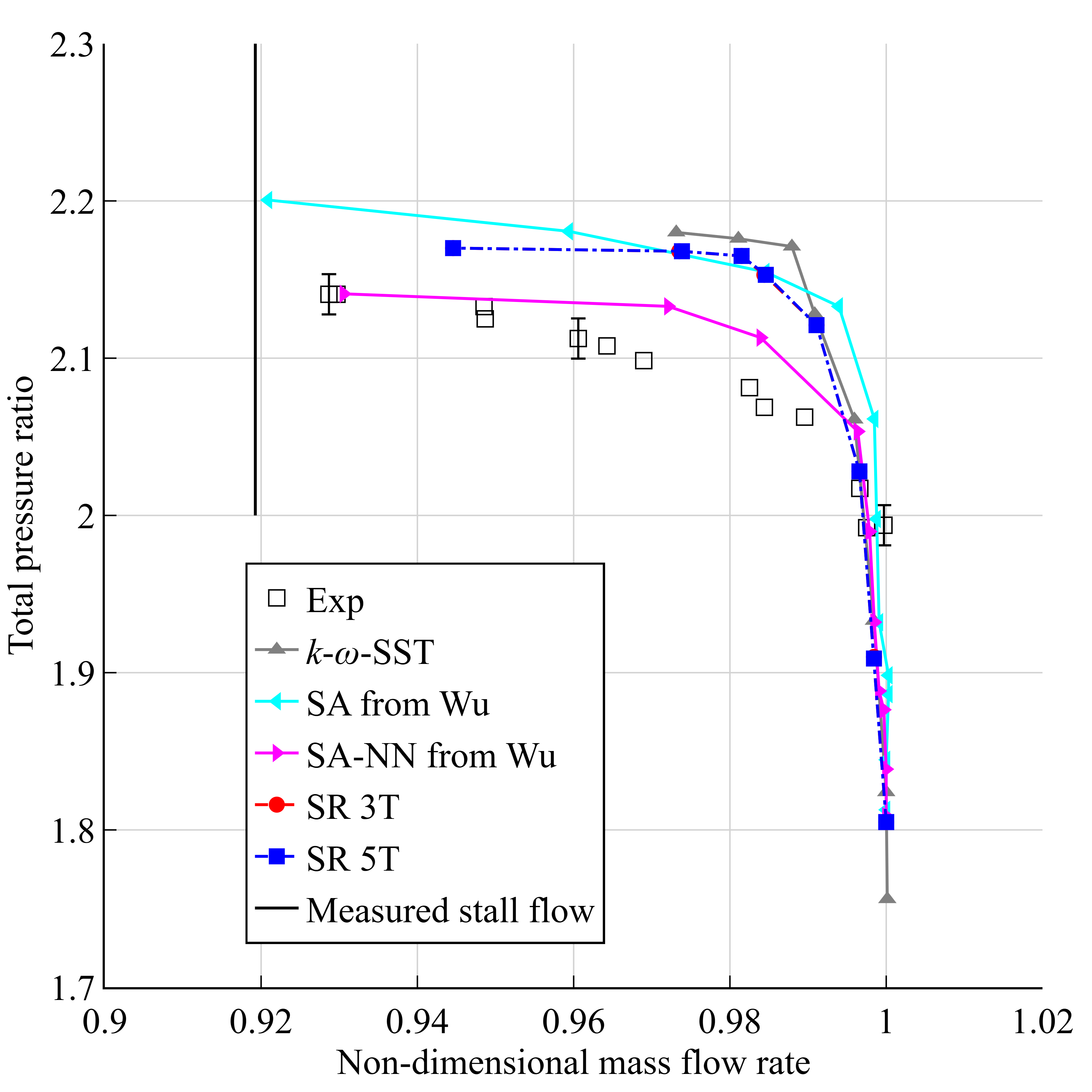}
  \centerline{(a)}
\end{minipage}
\hfill
\begin{minipage}{0.48\textwidth}
  \centering
  \includegraphics[width=\textwidth]{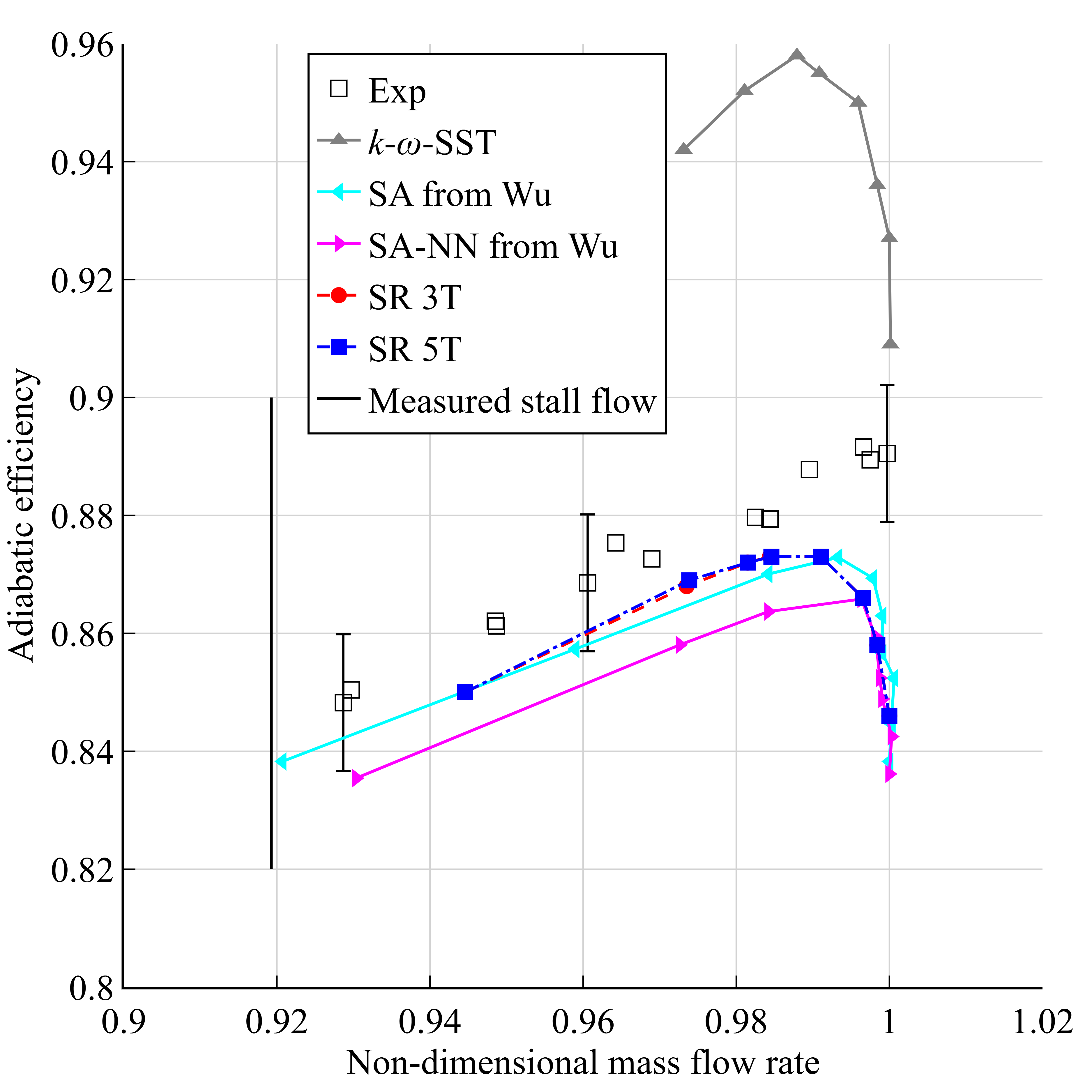}
  \centerline{(b)}
\end{minipage}
\caption{Overall performance of NASA Rotor 37. Plane (a) represents the total pressure ratio. Plane (b) represents the adiabatic efficiency.}
\label{Overall_performance}
\end{figure}

Figure \ref{Overall_performance} presents the overall performance of NASA Rotor 37. Panel (a) shows the total pressure ratio, while panel (b) displays the adiabatic efficiency. The results obtained using the Spalart-Allmaras (SA) model and the Spalart-Allmaras neural network (SA-NN) model are taken from \citet{wu_explainability_2023}, who developed a neural network-based turbulence model for simulating NASA Rotor 37 flows by directly utilizing experimental data from NASA Rotor 37 as training data. Their work provides a suitable benchmark for comparison with our results. Regarding the total pressure ratio, our SR 3T and SR 5T models yield similar predictions and remain stable down to approximately 0.94 non-dimensional mass flow rate. In contrast, the baseline $ k$-$\omega$-SST model can only achieve approximately 0.97. The symbolic regression turbulence models predict slightly lower total pressure ratios and show some improvement over the baseline $k$-$\omega$-SST model. Although the SA-NN model from \citet{wu_explainability_2023} achieves more accurate predictions, it should be noted that their model was trained directly on NASA Rotor 37 experimental data. In contrast, our models were trained exclusively on DNS results from periodic hill flows. The SA-NN model extends to a non-dimensional mass flow rate of 0.93, outperforming our models in this regard, primarily because their baseline SA model already exhibits strong performance in this regime. Regarding adiabatic efficiency, the SR 3T and SR 5T models again produce similar results. Notably, our symbolic regression models yield substantially improved predictions compared to the baseline $k$-$\omega$-SST model, with the results falling within the experimental uncertainty envelope over most of the operating range. While the baseline SA model also produces reasonably accurate predictions, the SA-NN model performs worse not only compared to our symbolic regression results but also relative to its own baseline SA model. It is worth emphasizing that the SA-NN model was trained on NASA Rotor 37 experimental data, which makes this outcome particularly noteworthy.

\begin{figure}
\centering \includegraphics[width=0.6\textwidth]{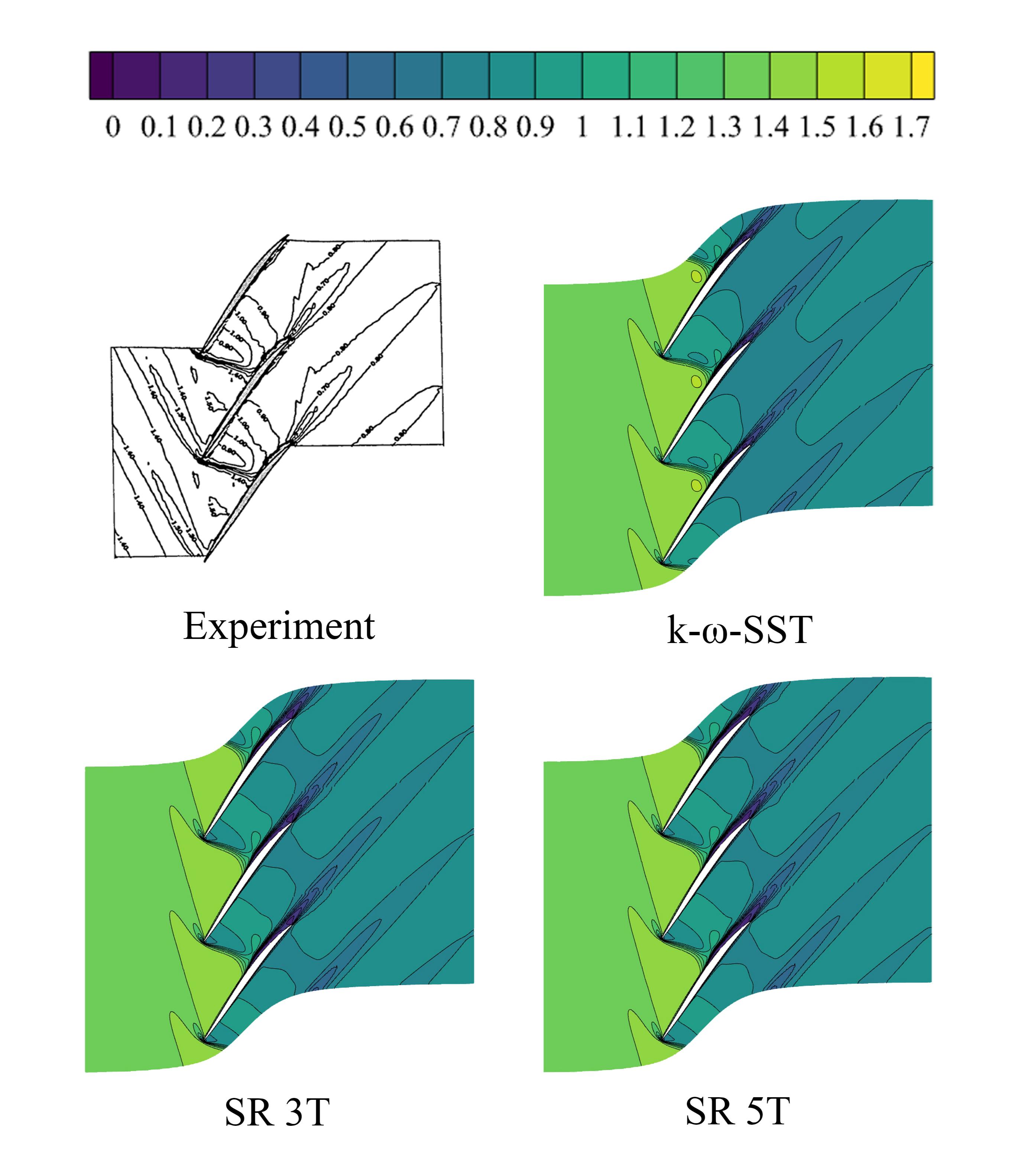} \caption{Relative Mach number contours at 70\% span of NASA Rotor 37 under 98\% measured choke flow condition.}\label{Relative Mach number contours}
\end{figure}

Figure \ref{Relative Mach number contours} presents the relative Mach number contours at 70\% span of NASA Rotor 37 under the 98\% measured choke flow condition. The results indicate that both the proposed symbolic regression models and the baseline $k$-$\omega$-SST model yield satisfactory predictions. Notably, the symbolic regression turbulence models accurately capture flow characteristics in the vicinity of the shock wave, despite being trained exclusively on incompressible flow data.

\begin{figure}
\centering \includegraphics[width=1.0\textwidth]{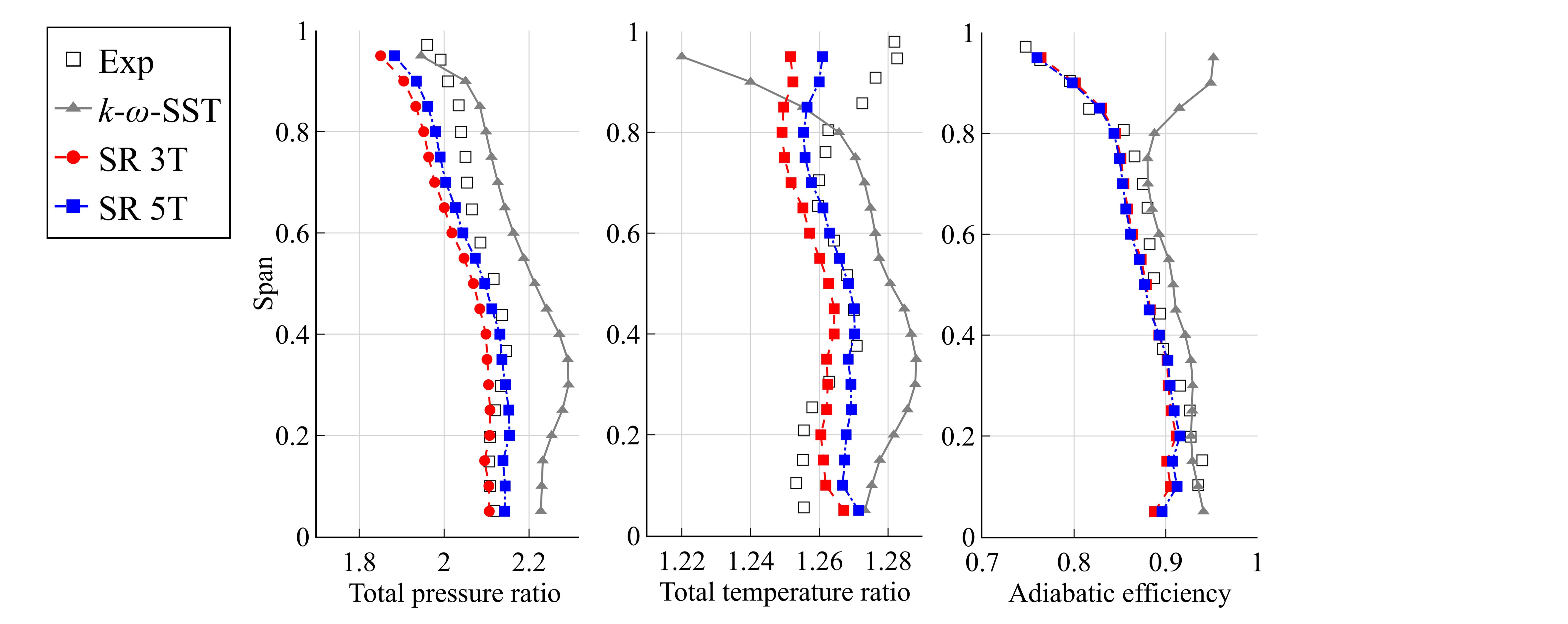} \caption{Radius distribution of total pressure ratio, total temperature ratio, and adiabatic efficiency on 98\% measured chock flow condition.}\label{Radius_distribution}
\end{figure}

Figure \ref{Radius_distribution} presents the radial distributions of the total pressure ratio, total temperature ratio, and adiabatic efficiency under the 98\% measured choke flow condition. Although the SR 3T and SR 5T models exhibit only minor differences in overall performance, as shown in figure \ref{Overall_performance}, noticeable discrepancies can be observed in the radial distributions of the total pressure ratio and total temperature ratio. For the total pressure ratio, both the SR 3T and SR 5T models show good agreement with the baseline $k$-$\omega$-SST model for spanwise locations below 0.4. In contrast, larger deviations are observed for spans above 0.6. For the total temperature ratio, the SR 5T model performs particularly well over the spanwise range 0.4-0.6 compared with the SR 3T and baseline $k$-$\omega$-SST models. Moreover, both the SR 3T and SR 5T models predict a curve shape that is in better agreement with the experimental results than that of the baseline $k$-$\omega$-SST model. For adiabatic efficiency, the SR 3T and SR 5T models provide accurate predictions in regions where the span exceeds 0.2.

Table \ref{Table: Radius_distribution_RE} summarizes the relative errors of the radial total pressure ratio, radial total temperature ratio, and radial adiabatic efficiency under the 98\% measured choked mass flow condition. The results indicate that the SR 5T model generally exhibits higher accuracy than both the baseline $k$-$\omega$-SST model and the SR 3T model. Specifically, the SR 5T model reduces the relative error in the total pressure ratio by 2.75\% and 0.69\% compared with the $k$-$\omega$-SST and SR 3T models, respectively. In terms of the total temperature ratio, the SR 5T model reduces the relative error by 1.02\% and 0.201\% relative to the $k$-$\omega$-SST and SR 3T models, respectively. Similarly, the relative error of the adiabatic efficiency is reduced by 3.76\% and 0.160\% when using the SR 5T model compared with the $k$-$\omega$-SST and SR 3T models, respectively.

\begin{table}
    \centering
    \begin{tabular}{cccc}
        Quantity & $k$-$\omega$-SST & SR 3T & SR 5T  \\ [3pt]
        Total pressure ratio & 4.69\% & 2.63\% & 1.94\%
 \\ 
        Total temperature ratio & 1.67\% & 0.847\%
 & 0.646\%
 \\ 
        Adiabatic efficiency  & 5.50\% & 1.90\%
 & 1.74\%
 \\ 
    \end{tabular}
     \caption{A summary of the relative errors of the radial total pressure ratio, radial total temperature ratio, and radial adiabatic efficiency under the 98\% measured choked mass flow condition.}
    \label{Table: Radius_distribution_RE}
\end{table}

\begin{figure}
\centering \includegraphics[width=1.0\textwidth]{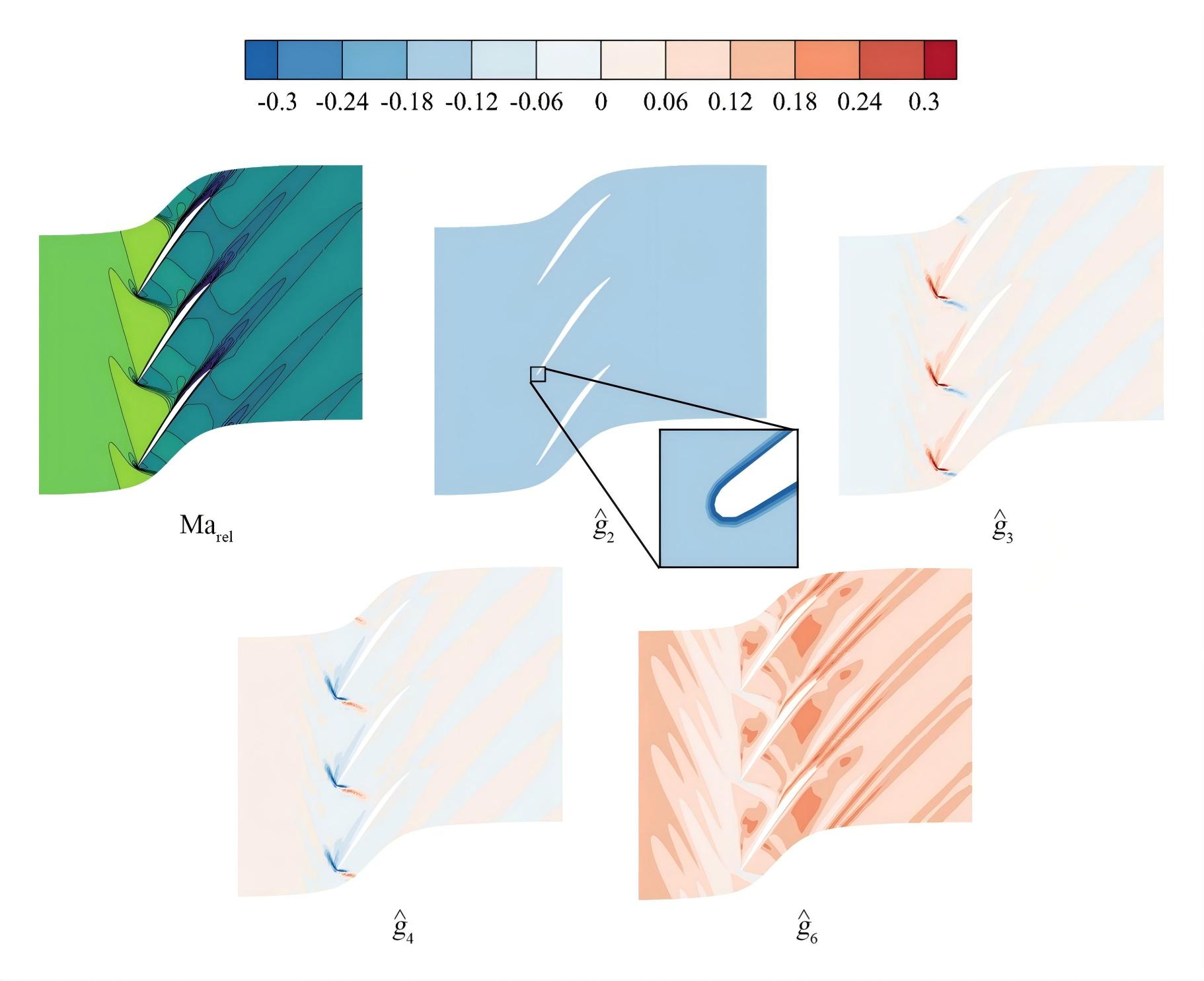} \caption{Contour plot of normalized tensor basis coefficients at 70\% span of NASA Rotor 37 under 98\% measured choke flow condition. The relative Mach number contour plot is identical to that in figure \ref{Relative Mach number contours}.}\label{Contour plot of normalized tensor basis coefficients NASA}
\end{figure}

Figure \ref{Contour plot of normalized tensor basis coefficients NASA} presents contour plots of the normalized tensor basis coefficients at 70\% span of NASA Rotor 37 under the 98\% measured choke flow condition. The relative Mach number contour is identical to that shown in figure \ref{Relative Mach number contours}. For $\hat{g}_2$, the regions of high absolute values are primarily concentrated in the near-wall region, indicating that the term $\hat{g}_2 \mathsfbi{\hat{T}}_2$ is predominantly activated in the near-wall regime. For $\hat{g}_3$ and $\hat{g}_4$, regions of high absolute values are observed in the vicinity of the shock wave, suggesting that the terms $\hat{g}_3 \mathsfbi{\hat{T}}_3$ and $\hat{g}_4 \mathsfbi{\hat{T}}_4$ are mainly activated in the shock-wave-dominated regime. For $\hat{g}_6$, high absolute values appear in the inlet region, within the blade passage, and in the wake region, indicating that the term $\hat{g}_6 \mathsfbi{\hat{T}}_6$ is primarily activated in these flow regimes.

\begin{figure}
\centering \includegraphics[width=1.0\textwidth]{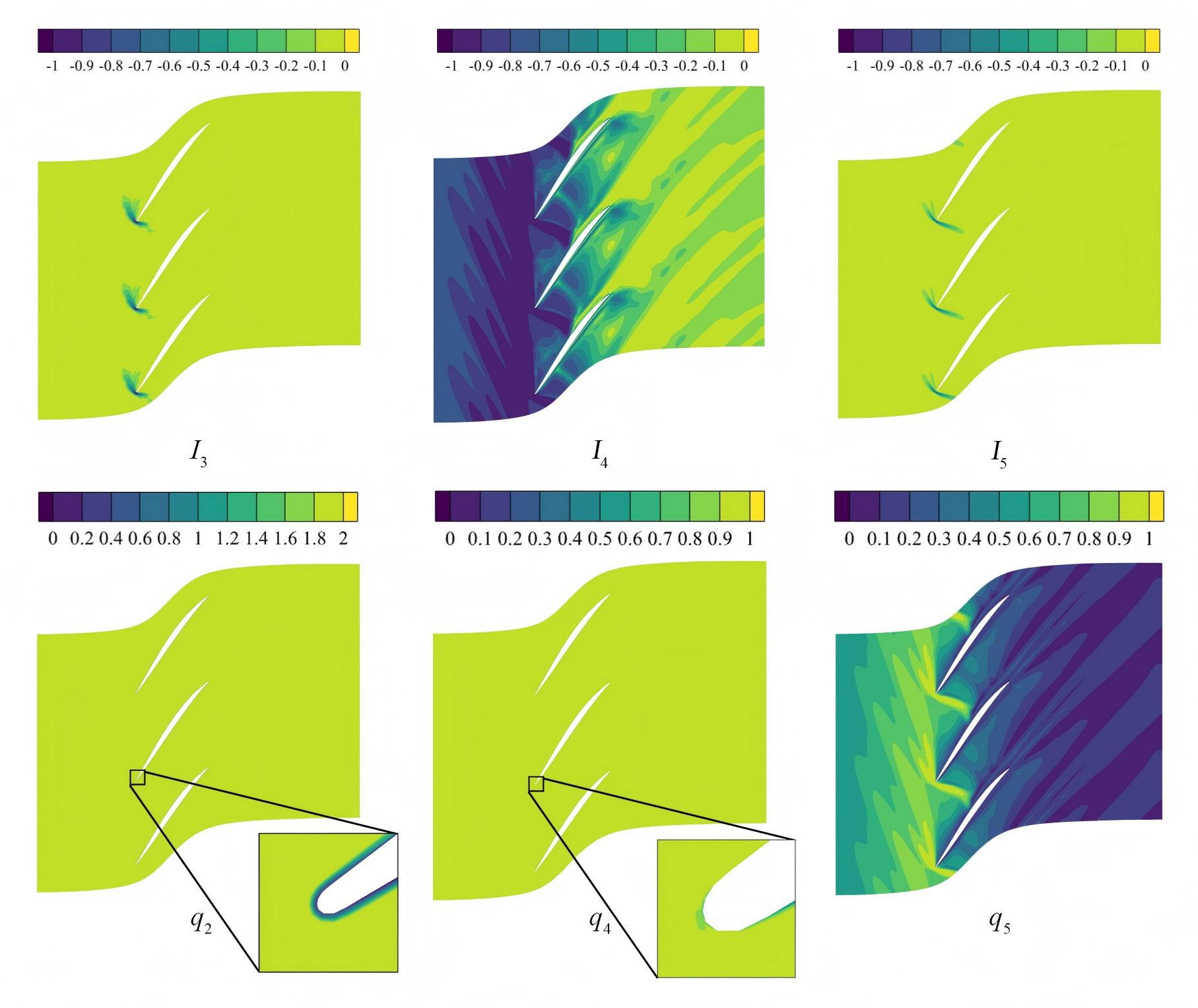} \caption{Contour plot of features at 70\% span of NASA Rotor 37 under 98\% measured choke flow condition.}\label{Contour plot of features NASA}
\end{figure}

Figure \ref{Contour plot of features NASA} presents contour plots of the features at 70\% span of NASA Rotor 37 under the 98\% measured choke flow condition. For $I_3$ and $I_5$, the values are nearly zero over most regions, except in the vicinity of the blade leading edge. For $I_4$, low-value regions are mainly located upstream, whereas high-value regions are predominantly distributed downstream. For $q_2$ and $q_4$, most low-value regions occur in the near-wall regime, while outside this regime the values generally approach their maximum. For $q_5$, high-value regions are primarily located upstream, whereas low-value regions are mainly distributed downstream. Based on these results and Fig. \ref{Contour plot of normalized tensor basis coefficients NASA}, the SR 5T model can be further analyzed. The term $\hat{g}_2 \mathsfbi{\hat{T}}_2$ is strongly activated when $q_2$ approaches zero. The terms $\hat{g}_3 \mathsfbi{\hat{T}}_3$ and $\hat{g}_4 \mathsfbi{\hat{T}}_4$ are activated in the vicinity of the shock-wave regime. For the term $\hat{g}_6 \mathsfbi{\hat{T}}_6$, it is mainly inactivated in the near-shock-wave regime (which is larger than the activated regimes of $\hat{g}_3 \mathsfbi{\hat{T}}_3$ and $\hat{g}_4 \mathsfbi{\hat{T}}_4$) and in the downstream region of the blades far away from blade wakes, while it is mainly activated in the blade-wake regime and within the blade passage. In summary, the SR 5T model differs from the classic Boussinesq hypothesis mainly in the near-wall region, the shock-wave region, the blade passage regime, and the blade wakes regime. It nearly restores the classic Boussinesq hypothesis in the blade-upstream region, where the norm of the turbulent kinetic energy gradient is much larger than the characteristic acceleration of turbulent fluctuations (corresponding to $I_4 + q_4 = 0$), and in the downstream region of the blades, far from blade wakes. The latter region is dominated by decaying turbulence with low shear and near isotropy, for which the classic Boussinesq hypothesis is already accurate. Therefore, this regime is expected to recover the classic Boussinesq hypothesis based on physical understanding. However, the blade-upstream region, where the turbulent kinetic energy gradient is strong, should not be expected to recover the classic Boussinesq hypothesis, based on physical considerations. We believe this behavior may be due to overfitting during training of the SR 5T model based on periodic hill flows.

\section{Discussion}\label{sec:Discussion}

Although the training data, comprising 3D slices of periodic hill flow DNS results, are not strictly two-dimensional, the RANS simulations of periodic hill flows presented in this article are entirely two-dimensional, rendering only three basis tensors linearly independent \citep{pope_more_1975, gatski_nonlinear_2000}. Nevertheless, the results demonstrate that the SR 5T model yields improved accuracy compared to the baseline $k$-$\omega$-SST model. In contrast, the SR 3T model provides negligible improvement and, in some cases, degrades the prediction quality. Based on the analysis presented in figure \ref{Contour plot of normalized tensor basis coefficients}, this discrepancy is primarily attributable to the contribution of the $\hat{g}_6 \mathsfbi{\hat{T}}_6$ term. Under two-dimensional conditions, $\mathsfbi{\hat{T}}_6$ becomes linearly dependent on $\mathsfbi{\hat{T}}_1$. However, direct symbolic regression modeling of $\hat{g}_1$ tends to prevent the model from recovering the baseline Boussinesq hypothesis. This consideration motivates the adoption of the SR 5T model in the present work and accounts for its superior accuracy.

\section{Conclusion}\label{sec:Conclusion}

In this study, we propose a few-shot, physically restorable symbolic regression turbulence model based on the normalized general effective-viscosity hypothesis. The model demonstrates good performance across several benchmark flow cases.

Specifically, we first propose a definition for few-shot data-driven turbulence models. The training is conducted exclusively on a three-dimensional slice extracted from the DNS data of periodic hill flows. Two symbolic regression models are developed: one incorporating three basis tensors that are linearly independent for two-dimensional flows, designated as the SR 3T model, and another incorporating five basis tensors, designated as the SR 5T model. A physical analysis of the symbolic regression results obtained from these data-driven turbulence models is subsequently presented. Subsequently, we employed four test cases to evaluate the accuracy of the symbolic regression turbulence models.

The first test case is the periodic hill flow, which was not included in the training data. Although the prior results for the periodic hill flow are not particularly promising, the posterior results demonstrate that the symbolic regression models have captured essential physical characteristics. As shown, the SR 5T model outperforms both the baseline $k$-$\omega$-SST model and the SR 3T model. The primary deviation from the baseline Boussinesq hypothesis occurs in regions characterized by low wall-distance-based Reynolds numbers and high mean strain rates. Further analysis reveals that the improved predictive accuracy for the periodic hill flow is primarily due to the term $\hat{g}_6 \mathsfbi{\hat{T}}_6$.

The second test case is the zero pressure gradient flat plate flow. The SR 5T model generally preserves the baseline $k$-$\omega$-SST model's accuracy. We observe that when the mean rotation time scale is significantly larger than the turbulence time scale, and the molecular viscous stress intensity dominates over the Reynolds stress intensity (e.g., in the viscous sublayer), the SR 5T model nearly reverts to the baseline Boussinesq hypothesis.

The third test case involves flow over an NACA0012 airfoil. The pressure coefficient distributions at angles of attack of $0^{\circ}$, $10^{\circ}$, and $15^{\circ}$ are nearly identical for the baseline $k$-$\omega$-SST model and the symbolic regression models, with all approaches yielding satisfactory predictions. Regarding the lift curve results, the SR 5T model performs best in the pre-stall regime. In contrast, the SR 3T model demonstrates superior accuracy in the post-stall regime.

The fourth test case examines the NASA Rotor 37 transonic axial compressor flows. The SR 5T model generally outperforms the baseline $k$-$\omega$-SST model and achieves higher accuracy in predicting adiabatic efficiency compared to a neural network-based turbulence model trained directly on experimental data from the NASA Rotor 37. The SR 5T model deviates from the baseline Boussinesq hypothesis primarily in the near-wall region, the shock-wave region, the blade passage, and the blade wake, while reverting to the baseline $k$-$\omega$-SST model in certain flow regimes.

In summary, our symbolic regression model is trained solely on periodic hill flows and generalizes well to engineering‑level turbulent flows. Moreover, in certain flow regimes, the symbolic regression turbulence model can physically recover the baseline $k$-$\omega$‑SST model behavior.

\backsection[Acknowledgements]{The authors thank all the reviewers and editors who contributed to improving the quality of this study.}

\backsection[Funding]{This research was supported by the National Science and Technology Major Project of China (J2019-II-00050025), the National Natural Science Foundation of China (No. 52406057), Research Grants Council of Hong Kong (No. CityU 21206123), and the National Natural Science Foundation of Guangdong Province (No. 2514050003981).}

\backsection[Declaration of interests]{The authors report no conflict of interest.}

\backsection[Data availability statement]{The data that support the findings of this study are available from the corresponding author upon reasonable request.}

\backsection[Author ORCIDs]{Z. Ji, https://orcid.org/0009-0004-0032-6251; P. Duan, https://orcid.org/0000-0002-3501-8855; G. Du, https://orcid.org/0009-0000-4296-3807}

\backsection[Author contributions]{Z. Ji: Conceptualization (lead); Data curation (lead); Formal analysis (lead); Investigation (lead); Methodology (lead); Software (lead); Validation (lead); Visualization (lead); Writing - original draft (lead); Writing - review \& editing (equal). P. Duan: Supervision (equal); Writing - review \& editing (equal). G. Du: Funding acquisition (lead); Supervision (equal); Writing - review \& editing (equal).}

\appendix

\bibliographystyle{jfm}
\bibliography{jfm}


\end{document}